\documentclass{aa}  
\usepackage[]{natbib}
\usepackage{graphicx}
\usepackage{txfonts}
\usepackage{colortbl}
\usepackage[dvipsnames]{xcolor}
\usepackage{epstopdf}
\usepackage{hyperref}
\usepackage{lscape}
\usepackage[switch]{lineno}
\usepackage{placeins}
\usepackage{stfloats}

\usepackage{cancel} 
\usepackage{soul}

\usepackage{ulem}

\begin{document} 

\title{The advanced evolution of massive stars}

   \authorrunning{T. Dumont et al.}
   
   \titlerunning{The advanced evolution of massive stars: I. New reaction rates for carbon and oxygen nuclear reactions} 
   
   \subtitle{I. New reaction rates for carbon and oxygen nuclear reactions}

   \author{T. Dumont
          \inst{1} \fnmsep\thanks{e-mail: thibaut.dumont@iphc.cnrs.fr},
          A. Bonhomme \inst{1}, A. Griffiths \inst{2}, A. Choplin \inst{3}, M. A. Aloy \inst{2,4}, G. Meynet \inst{5}, K. Godbey \inst{6}, C. Simenel \inst{7}, G. Scamps \inst{8},  F. Castillo \inst{1}, A. Cosoli-Ortega \inst{1}, and S. Courtin \inst{1,9}\\
          }

   \institute{University of Strasbourg, CNRS, IPHC UMR 7178, F-67000 Strasbourg, France
         \and Departament d’Astonomia i Astrofísica, Universitat de València, C/Dr. Moliner, 50, E-46100 Burjassot (València), Spain
         \and Institut d'Astronomie et d'Astrophysique, Université Libre de Bruxelles, CP 226, 1050, Brussels, Belgium
         \and Observatori Astronòmic, Universitat de València, 46980 Paterna, Spain
         \and Department of Astronomy, University of Geneva, Chemin Pegasi 51, 1290 Versoix, Switzerland
         \and Facility for Rare Isotope Beams, Michigan State University, East Lansing, Michigan 48824, USA.
         \and Department of Fundamental and Theoretical Physics, and Department of Nuclear Physics and Accelerator Applications, Research School of Physics, Australian National University, Canberra ACT 2600, Australia
         \and Laboratoire des 2 Infinis - Toulouse (L2IT-IN2P3), Université de Toulouse, CNRS, UPS, F-31062 Toulouse Cedex 9, France.
         \and University of Strasbourg, Institute of Advanced Studies (USIAS), Strasbourg, France
             }

   \date{}

 
  \abstract
   {The nuclear rates for reactions involving $^{12}$C and $^{16}$O are key to computing the energy release and nucleosynthesis evolution of massive stars during their advanced burning phases. Ultimately, these burning rates shape the stellar structure and evolution and influence the nature of the compact objects produced at the end of the stellar life.
   }
   {We explore the implications of new nuclear reaction rates from both experimental and theoretical studies for $\rm ^{12}C(\alpha,\gamma)^{16}\!O$, $\rm ^ {12}C+^{12}\!C$, $\rm ^{12}C+^{16}\!O$, and $\rm ^{16}O+^{16}\!O$ reactions for massive stars. Our goal is to investigate how the chemical structure and nucleosynthesis evolve from the He-exhaustion stage to the O-burning phase and how these processes influence the ultimate stellar fate. 
   }
   {We computed rotating and non-rotating models for stars of different masses at solar metallicity. We used the stellar evolution code GENEC, which includes a large network of nuclear reactions and isotopes involved in advanced phases, as well as updated rates for $\rm ^ {12}C(\alpha,\gamma)^{16}\!O$. For the three fusion reactions involving $^{12}$C and $^{16}$O, we considered new rates following a data-driven fusion suppression scenario (hereafter HIN(RES)) and new theoretical rates obtained with time-dependent Hartree-Fock (TDHF) calculations.
   }
   {The updated $\rm ^ {12}C(\alpha,\gamma)^{16}\!O$ rates mainly impact the chemical structure evolution changing the $^{12}$C/$^{16}$O ratio at He-exhaustion and have little effect on the CO core mass. This variation in the $^{12}$C/$^{16}$O ratio is in some cases critical for predicting the final fate of the model, which is very sensitive to $^{12}$C abundance, and in particular the 20~M$_{\odot}$ remnant may change from a black hole to a neutron star. The He-burning (C-burning) lifetime is also decreased (increased) by about $-2 \%$ ($+15 \%$). The combined new rates for $\rm ^ {12}C+^{12}\!C$ and $\rm ^ {16}O+^{16}\!O$ fusion reactions according to the HIN(RES) model lead to shorter C- and O-burning lifetimes by $\approx$ -10\%, and -50\%, respectively, and shift the ignition conditions to higher temperatures and densities. In contrast, the theoretical TDHF rates primarily affect C-burning, increasing its duration by about 30\% and lowering the ignition temperature. These changes modify the chemical structure of the core, the size and duration of C-burning shells, and hence their compactness. They also impact the central and shell nucleosynthesis (by $\rm \pm$ 1~dex and by factors of $\pm$ 2-10, respectively), while $\rm ^ {12}C+^{16}\!O$ reaction rates variations remain the least important. 
   }
   {The present work shows that accurate reaction rates for key processes in massive star evolution and nucleosynthesis drive significant changes in stellar burning lifetimes, chemical evolution, and stellar fate. The multiple and cumulative consequences of these changes are significant and should not be neglected. In addition, discrepancies between experimental and theoretical rates introduce uncertainties in model predictions, influencing both the internal structure and the composition of the supernova ejecta.
   }


   \maketitle
   
%

\section{Introduction}
\label{section:introduction}

Massive stars are keystones of the chemical evolution of the Universe, synthesising heavy elements in their interiors and enriching their surroundings through Type II supernovae explosions \citep[e.g.][]{Woosley2002}. The supernova ejecta composition reflects the cumulative effects of successive nuclear burning stages and extensive nucleosynthesis during massive star evolution \citep[e.g.][]{Bennett2012,Chieffi2013}. A detailed understanding of the stellar evolution phases from hydrogen to silicon burning and their role in driving nucleosynthesis remain incomplete, especially for the most advanced phases, i.e. beyond  He-exhaustion. The specific density and temperature conditions reached in these phases enable a broad range of nuclear reactions to occur, including s-process nucleosynthesis, contributing to heavy element formation. These reactions also influence the stellar fate by determining the remnant type and mass as well as the composition of the supernova ejecta \citep[e.g.][]{Chieffi2020}. Only the most energetic reactions involving the most abundant elements significantly drive stellar evolution and impact the chemical structure. Among these, nuclear reactions involving carbon and oxygen are key to massive star evolution \citep[e.g.][]{Woosley1986}. Ongoing experimental and theoretical efforts aim to determine the rates of crucial reactions at the relatively low energies of astrophysical interest with improved accuracy and reliability \citep[e.g.][]{2018PrPNP..98...55B}. 

Four reactions involving carbon and oxygen are of particular interest in the context of advanced evolution. First, the $\rm ^{12}C(\alpha,\gamma)^{16}O$ reaction, primarily active during the latter part of He-burning phase, governs the $^{12}$C/$^{16}$O ratio in stars and potentially affects the CO core mass at the end of the He-burning phase. In particular, \citet{deBoer2017} have shown that lower reaction rates lead to a higher $^{12}$C/$^{16}$O ratio with consequences on stellar evolution and structure, eventually influencing the explosiveness \citep[see also e.g.][]{Sukhbold2020,Xin2025}. The $^{12}$C/$^{16}$O ratio is ultimately propagated throughout the Universe when a massive star explodes, ejecting the carbon and oxygen from its outer layers, which then contribute to the formation of new stars. Second, the fusion reaction $\rm ^ {12}C+^{12}\!C$, the most important reaction during the C-burning phase, impacts both the nucleosynthesis and the final fate of stars \citep[e.g.][]{Bennett2012,Pignatari2013,Frischknecht2016,Limongi2018,Chieffi2021,Monpribat2022,Dumont2024}. Third, the fusion reaction $\rm ^ {16}O+^{16}\!O$ is the main reaction during the O-burning phase. Finally, the fusion reaction $\rm ^ {12}C+^{16}\!O$ is expected to occur at the transition between carbon and oxygen fusion, mainly during shell burning. This reaction likely contributes to heavy element production from the end of the core C-burning phase to the core O-burning phase.

Stellar evolution models rely on nuclear physics, with both the choice of isotopes and the adopted reaction network strongly influencing stellar evolution and nucleosynthesis. Nuclear reaction rates are obtained from experimental cross-section measurements and theoretical predictions, which are both subject to uncertainties. At the low energies typical of stellar interiors (a few mega electron-volts), nuclear reaction cross-sections are extremely small, making precise and reliable rate measurements particularly challenging. Consequently, variations in these rates propagate into uncertainties in stellar model predictions. While widely used nuclear rate compilations include \citet[][hereafter CF88]{Caughlan1988}, \citet[][Nacre]{Angulo1999}, and \citet[][NacreII]{Xu2013a}, newer rates are available, offering the potential to improve our modelling of stellar nuclear physics processes. The aim of the this paper is to explore the newly available reaction rates with regard to their potential impact on the advanced stages of stellar evolution, nucleosynthesis, and the ultimate fate of massive stars, with a particular focus on chemical abundances and stellar structure.
 
In a previous paper, \citet[][hereafter D24]{Dumont2024} examined the impact of updated nuclear reaction rates for $\rm ^ {12}C+^{12}\!C$ during the core C-burning phase. Here, we extend this analysis by incorporating revised nuclear rates for the $\rm ^{12}C(\alpha,\gamma)^{16}O$ reaction. We also expand our investigation to include the O-burning phase, exploring the consequences of adopting new reaction rates based on the Hindrance scenario or on theoretical nuclear model predictions for the $\rm ^ {12}C+^{16}\!O$ and $\rm ^ {16}O+^{16}\!O$ reactions\footnote{This work is part of the STELLA-CarbOx project, which aims to measure cross-sections for $\rm ^ {12}C+^{16}\!O$ and $\rm ^ {16}O+^{16}\!O$ reactions.}. 

The outline of this paper is as follows. In Sect.~\ref{GENEC}, we describe the input physics included in the Geneva stellar evolution code (GENEC). In Sect.~\ref{sect:nucl}, we describe the nuclear rates that we are testing in our models. In Sect.~\ref{sect:class}, we present the impact on stellar evolution from carbon burning up until oxygen burning. 
Sect.~\ref{sect:nucleo} presents the resulting heavy-element nucleosynthesis, with particular emphasis on the effects on the s-process. In Sect.~\ref{sect:remnant}, we asses the implications of our models for the final fate of massive stars. Finally, we discuss our results and conclude in Sect~\ref{sect:CONCLUSION}.

\section{The stellar evolution code}
\label{GENEC}
We modelled the evolution of massive stars using the stellar evolution code GENEC as described in \citet{Eggenberger2008,Ekstrom2012,Griffiths2025} and D24. We computed both rotating and non-rotating models at solar metallicity (Z=0.014) for three different initial masses: 15, 17, and 20~M$_{\odot}$. Each model was evolved from the zero age main-sequence (ZAMS) to O-exhaustion, defined as when the oxygen core abundance $X_{^{16}O}$ drops below $10^{-3}$, unless otherwise stated.

The latest version of the GENEC code was developed to better capture the advanced phases of massive star evolution \citep[][]{Griffiths2025}. In this version, radiative opacities are generated within the OPAL opacity tables from \citet{Iglesias1996} for temperatures lower than 1~Gk. At higher temperatures, the estimates from \cite{Poutanen2017} are used. The equation of state (EoS) combines a low temperature and density regime, which includes partial ionisation effects, with the fully ionised plasma EoS of \citet{Timmes2000}. These two regimes are matched at conditions log(T)>7.55 and log($\rho$)>2.8. For a typical 25 M$_{\odot}$ star, this corresponds to a transition between the two EOSs in the core during the He-burning phase. We used the grey atmosphere approximation and treated convection according to mixing length theory, with  $\rm \alpha_{MLT} = 1.6$, calibrated to the Sun with the solar reference abundances from \citet{Asplund2005} and \citet{Cunha2006}, as in D24. We used the Schwarzschild criteria for convective stability and applied core overshooting for phases prior to C-ignition with an overshoot parameter $\rm d_{over}$ = 0.1 Hp, where Hp is the pressure scale height. In the present paper, we use the same formalism for rotation as described in D24 with $\rm V_{ini}/V_{crit}$ = 0.4 and follow the shellular rotation hypothesis by \citet{Zahn1992,Maeder1998,Mathis2004}. For the sake of comparison with D24, we do not take into account the action of magnetic fields. However, magnetic torques are drivers of angular momentum transport in stellar evolution. The  combined impact of the new rates and magnetic angular momentum transport will be the object of a future work.

\section{Nuclear reaction network and rates}
\label{sect:nucl}

\begin{table}[t]
    \centering
    \caption{Reference for reactions important during C-burning and O-burning phases.}
    \begin{tabular}{c|c|c}
    \hline \hline
    Reaction & D24 & This work \\
    \hline
    $\rm ^{12}C(\alpha,\gamma)^{16}O$ &  \cite{Kunz2002} & \cite{deBoer2017} \\
    $\rm ^{12}C+^{12}\!C$ &  M22$^\star$ & HIN(RES) or TDHF \\
    $\rm ^{12}C+^{16}\!O$ &  CF88 & HIN(RES) or TDHF \\
    $\rm ^{16}O+^{16}\!O$ &  CF88 & HIN(RES) or TDHF \\
    \hline
    \end{tabular}
    \label{tab:reactions_1}
    \tablefoot{M22: \citet{Monpribat2022}. TDHF: Time-dependent Hartree-Fock. CF88: \citet{Caughlan1988}. $^\star$ Recommended HinRes model from M22.}
\end{table}

The nuclear network used in GENEC during the advanced evolutionary phases has been updated from the version employed in D24 to include the primary reactions that drive the energy generation and chemical evolution during C-, Ne-, and O-burning. The current network, GeValNet48, comprises 48 species. It builds on the spine network GeValNet25 introduced in \citet{Griffiths2025}, with the addition of secondary alpha chains to enable the tracking of reactions such as $\rm ^{16}O(\rm ^{16}O,\rm p)\rm ^{31}P$. The full list of explicitly tracked isotopes is detailed in App.~\ref{Annexe:nucl_net}. Furthermore, we have updated the reaction rates of seven key nuclear reactions that significantly affect the most abundant elements and the neutron density following the recommendations of the Netgen database\footnote{http://www.astro.ulb.ac.be/Netgen/} and as identified by \citet{Choplin2018} (see Sect.~\ref{sec:neutron-reactions}). For reactions not directly involved in our study, rates were taken from the NACRE II database \citep{Xu2013a} or from CF88 when NACRE II data were not available. The main focus of the present study is to explore the impact and sensitivity of the measured or theoretically predicted reaction rates (listed in Tab.~\ref{tab:reactions_1}) for key nuclear reactions involving $\rm ^{12}C$ and $\rm ^{16}O$. These include the $\rm ^{12}C(\alpha,\gamma)^{16}O$ reaction, which plays a critical role during He-burning as well as the $\rm ^{12}C+^{12}\!C$, $\rm ^{12}C+^{16}\!O$, and $\rm ^{16}O+^{16}\!O$ fusion reactions that occur during the later stages of stellar evolution.

\subsection{$\rm ^{12}C(\alpha,\gamma)^{16}O$}
\label{sect:cago}

\begin{figure*}[t]
    \includegraphics [trim = 5mm 15mm 10mm 5mm,clip,width=90mm]{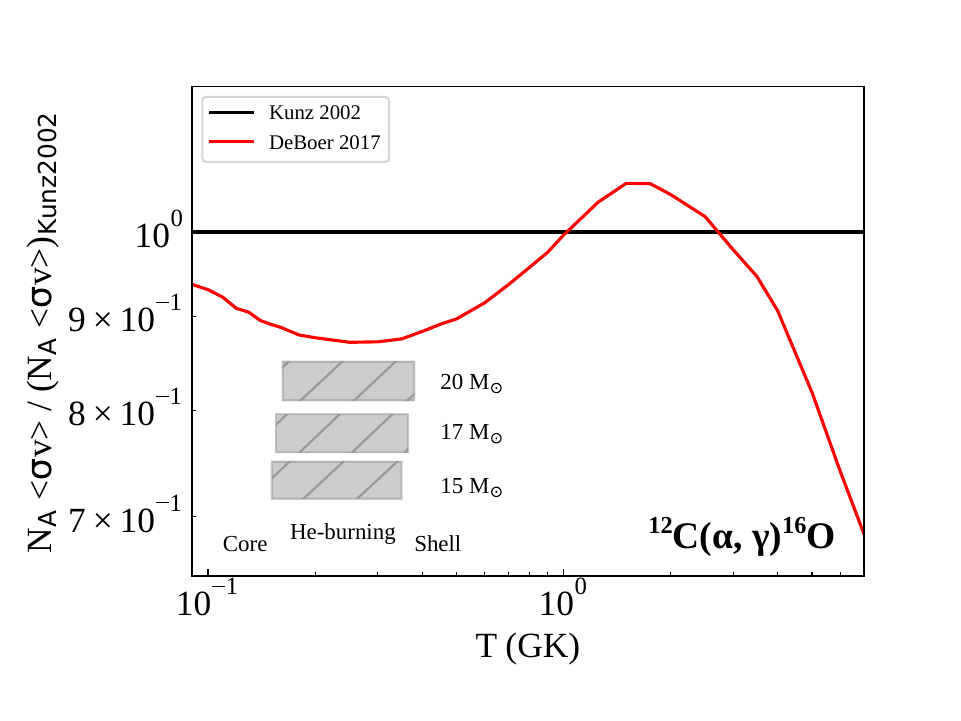}
    \includegraphics [trim = 5mm 15mm 10mm 5mm,clip,width=90mm]{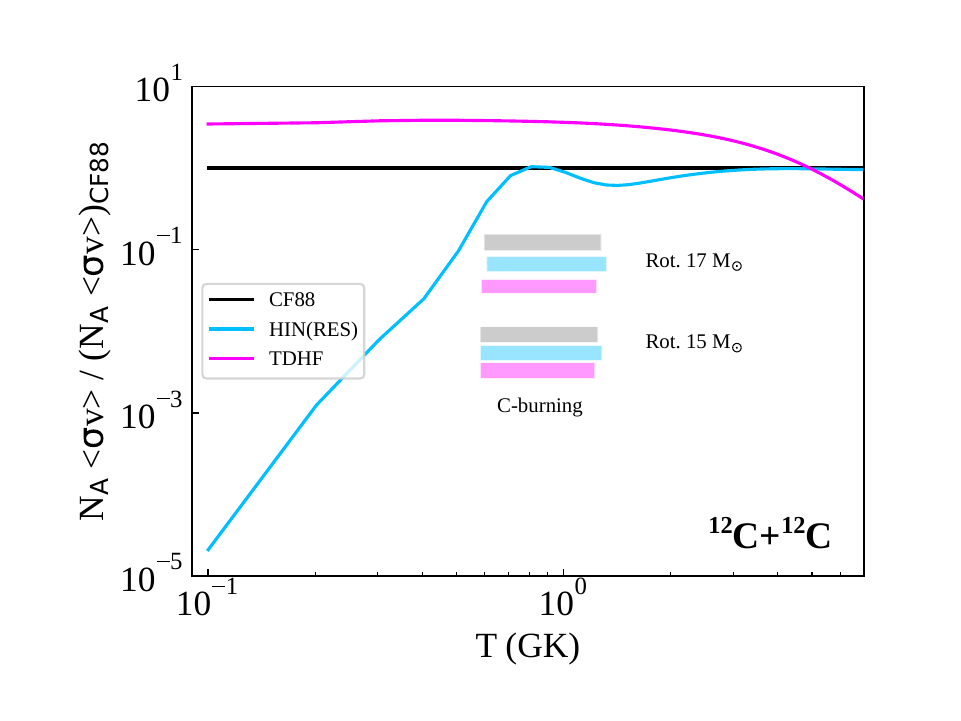}
    \includegraphics [trim = 5mm 8mm 10mm 10mm,clip,width=90mm]{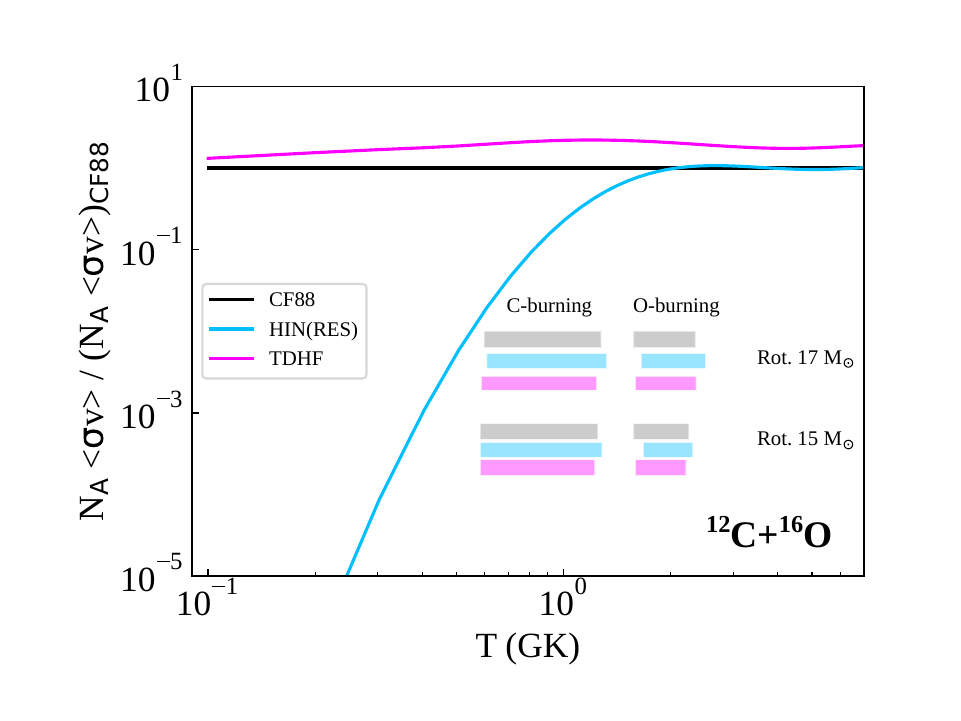}
    \hspace{2mm}
    \includegraphics [trim = 5mm 8mm 10mm 10mm,clip,width=90mm]{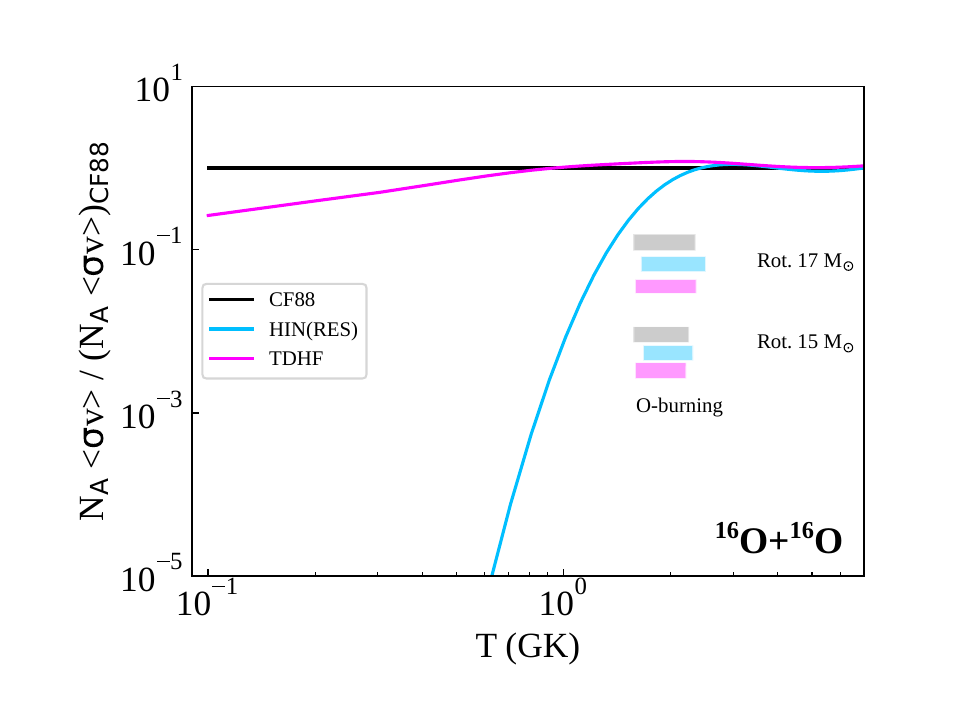}
    \caption{Reaction rates for the four main nuclear reactions involving $^{12}$C and $^{16}$O. Top left: Rates from \citet{deBoer2017} normalised to \citet{Kunz2002} reference for the $\rm ^{12}C(\alpha,\gamma)^{16}O$ reaction. The grey hatched areas indicate the temperature ranges where core and shell He-burning occur in the GENEC rotating model explored mass range. For the rest of the panels, nuclear reaction rates are taken from CF88 (black), HIN(RES (blue), and TDHF (magenta), all normalised to the CF88 rates ($N_A <\sigma v>_{CF88}$). Shaded areas (with corresponding colour codes) indicate the temperature ranges where core and shell C-burning and core O-burning take place in the 15 and 17~$M_{\odot}$ GENEC rotating models.
    }
    \label{fig:rates_multi}
 \end{figure*}

One of the most important nuclear reactions during stellar evolution is $\rm ^{12}C(\alpha,\gamma)^{16}O$. This reaction primarily occurs during the He-burning phase, in competition with the $3\alpha$ process, and drives the $\rm ^{12}C/^{16}O$ ratio at the end of this phase. In the present work, we adopt the updated reaction rates from \cite{deBoer2017} instead of those from \citet{Kunz2002}, which were used in D24. The rates provided by \cite{deBoer2017} are more accurate and have smaller uncertainties, as illustrated in their Fig.~29. Figure~\ref{fig:rates_multi} compares the two sets of rates and shows the predicted He-burning temperatures in both the core and the shell. In the region of interest, the \citet{deBoer2017} rates are systematically lower than those of \citet{Kunz2002}. This revision results in a slightly shorter He-burning lifetime (by $\sim$2\%)\footnote{He-burning lifetimes of roughly 1.30 Myr, 1.08, and 0.87 are obtained for the rotating 15, 17, and 20~M$_{\odot}$ models, respectively, compared to 1.31, 1.14, and 0.88 Myr.} 
, an approximately 15\% increase in the C-burning lifetime, shorter Ne- and O-burning lifetimes (according to internal report), and a higher $\rm ^{12}C/^{16}O$ ratio at the end of He-burning, along with a slightly hotter and denser core. 

According to \citet{Patton2020}, both the CO core mass and the central $\rm ^{12}C$ mass fraction at the end of He-burning influence the final fate of massive non-rotating stars. Variations in these parameters between models can thus affect the nature of the compact remnant produced at the end of stellar evolution. This has been recently confirmed, for example, by \citet{Xin2025} (see also references therein), who examined the impact of an arbitrary $\pm$3$\sigma$ variation in reaction rates, based on the \citet{Kunz2002} data, and showed, using a different methodology than the one employed here, that such variations influence both the mass and type of remnant. 

In our models, we find that the CO core mass at the end of He-burning is largely insensitive to changes in the $\rm ^{12}C(\alpha,\gamma)^{16}O$ reaction rates. For instance, we obtain M$_{\rm CO}\approx 2.35~M_\odot$, $2.92~M_\odot$, and 4.10~M$_\odot$ for our 15, 17, and 20~M$_\odot$ models, respectively.
 
However, the $\rm ^{12}C/^{16}O$ ratio is affected, increasing from $\rm ^{12}C/^{16}O\approx 0.55$ with the \citet{Kunz2002} rates to $\rm ^{12}C/^{16}O\approx 0.65$ with the updated rates from \cite{deBoer2017}, a relative increase of about 20\%. This is expected, as the lower $\rm ^{12}C(\alpha,\gamma)^{16}O$ rate leads to fewer $\alpha$ captures by $\rm ^{12}C$, resulting in more $\rm ^{12}C$ remaining and less $\rm ^{16}O$ being synthesised. The small change observed in the CO core mass is also consistent with expectations, as this quantity is primarily determined by the convective core mass during He-burning, which in turn depends on the total energy released by nuclear reactions. Since the $\rm ^{12}C(\alpha,\gamma)^{16}O$ reaction does not dominate the energy generation in this phase, the overall energy output remains nearly unchanged between the two models. We will further explore the implications for the final fate of these models in Sect.~\ref{sect:remnant}.

\subsection{Fusion reactions $\rm ^{12}C+^{12}\!C$, $\rm ^{12}C+^{16}\!O$ and $\rm ^{16}O+^{16}\!O$}

In \cite{Monpribat2022}, hereafter M22, and D24, it was shown that recent measurements for the $\rm ^{12}C+^{12}\!C$ reaction exhibit significant deviations from the commonly used CF88 rates. 
These updated rates account for a suppression trend (or \textit{hindrance}) of the fusion cross-section at low energies, and also incorporate a resonance at 2.14\,MeV, first observed by \citet{Spillane2007} and later supported by STELLA measurements \citep[][]{Fruet2020}. We previously demonstrated that the adoption of these new rates yields a shorter C-burning lifetime, a hotter and denser burning core, and a shift in the transition from a convective C-burning core to a radiative C-burning core to higher initial masses (approximately 29 and 25~M$_{\odot}$ for rotating and non-rotating models, respectively, compared to about 27 and 25~M$_{\odot}$ with the CF88 rates). Here, we extend the analysis beyond core C-burning to explore the impact of these updated rates on the C-shell burning phase, as well as on core Ne- and core O-burning phases. In these later stages, the $\rm ^{16}O+^{16}\!O$ fusion reaction becomes increasingly relevant, along with a potential contribution from the $\rm ^{12}C+^{16}\!O$ reaction. Given their similarity to the $\rm ^{12}C+^{12}\!C$ reaction, we adopt a consistent modelling approach for all three reactions and compare the resulting outcomes. We focus mainly on three representative stellar models, as described below. In the absence of experimental data within the relevant Gamow windows for these reactions, this study also serves as a preliminary sensitivity study aimed at guiding future experimental efforts.

As the first energetic reaction of interest for massive stars after He-burning, carbon fusion takes place for temperatures of at least 0.6-0.8~GK:
\begin{equation}
    \rm
     ^{12}C+^{12}\!C \xrightarrow{}^{24}\!Mg^* \left\{ 
     \begin{array}{lll}
                \xrightarrow{} \rm ^{20}Ne+\alpha & \text{(Q = 4.62 MeV)} \\
                \xrightarrow{} \rm ^{23}Na+p & \text{(Q = 2.24 MeV)} \\
                \xrightarrow{} \rm ^{23}Mg+n & \text{(Q = -2.62 MeV)}
    \end{array}
    \right\},
    \label{eq:channels1}
\end{equation}
where Q (or Q-value) refers to the energy released or consumed by the reaction.
For this reaction in particular, substantial experimental efforts and advances in measurement techniques have enabled direct measurements to probe increasingly lower energies. However, the available data are still suffering from significant uncertainties at the centre-of-mass at energies below 2\,MeV \citep[][]{Spillane2007,2018PhRvC..97a2801J,Fruet2020}. To overcome the challenges posed by low-energy measurements, alternative indirect methods, such as the so-called Trojan Horse Method (THM) \citep{Mukhamedzhanov2008}, have been developed. This technique provides indirect estimates of the $\rm ^ {12}C+^{12}\!C$ reaction cross-section, which are several orders of magnitude higher than those obtained via direct measurements \citep[][]{Tumino2018}. While these data have been employed in stellar evolution studies \citep{Chieffi2021,Xin2025}, their interpretation is remains contentious within the community \citep[see e.g.][]{Mukhamedzhanov2019}, further work is needed to reconcile the results of these two complementary approaches. In the present paper, we focus exclusively on the impact of the new rates obtained from direct measurements and theoretical nuclear model predictions, within the temperature range relevant for core and shell C-burning phases. 

At temperatures above 1~GK, during both hydrostatic and explosive C- and O-burning phases, when the $\rm ^ {12}C+^{12}\!C$ fusion reaction becomes subdominant, the following nuclear reaction becomes relevant:
\begin{equation}
    \rm
     ^{12}C+^{16}\!O \xrightarrow{}^{28}\!Si^* \left\{ 
     \begin{array}{lll}
                \xrightarrow{} \rm ^{24}Mg+\alpha & \text{(Q = 6.77 MeV)} \\
                \xrightarrow{} \rm ^{27}Al+p & \text{(Q = 5.17 MeV)} \\
                \xrightarrow{} \rm ^{27}Si+n & \text{(Q = -0.42 MeV)}
    \end{array}
    \right\},
    \label{eq:channels2}
\end{equation} 
Suitable conditions for reaction (\ref{eq:channels2}) occur when the $^{16}$O abundance is high enough (typically between core C-exhaustion and the onset of O-burning, as well as in the shell C-burning regions). Although this reaction has a lower probability to occur than reactions (\ref{eq:channels1}) and (\ref{eq:channels3}), its main impact lies not in the energy production but in its potential to influence nucleosynthesis yields during the C- and O-burning phases. Experimentally, this reaction has recently been revisited \citep[e.g.][]{Fang2017} at low energies of astrophysical relevance, enriching the dataset that has been available since the 70s. These new measurements provide modest evidence for the existence of resonances in the deep sub-Coulomb regime.

The last major reaction for which we vary the rates is the dominant nuclear process during the O-burning phase, occurring at temperatures of about 1.9~GK \citep[e.g.][]{Maeder2009}:
\begin{equation}
    \rm
     ^{16}O+^{16}\!O \xrightarrow{}^{32}S^* \left\{ 
     \begin{array}{lll}
                \xrightarrow{} \rm ^{28}Si+\alpha & \text{(Q = 9.59 MeV)} \\
                \xrightarrow{} \rm ^{31}P+p & \text{(Q = 7.68 MeV)} \\
                \xrightarrow{} \rm ^{31}S+n & \text{(Q = 1.45 MeV)}
    \end{array}
    \right\},
    \label{eq:channels3}
\end{equation}
Due to the experimental difficulties to measure this reaction accurately, mainly caused by the technical availability of oxygen targets and the large number of exit channels, only scarce and sometimes disagreeing measurements from the 70s and 80s are currently available \citep[][and see for instance Fig.~1 by \citealt{Simenel2013}]{Fernandez:1978qpm,PhysRevC.18.1688,1987PhRvC..35..591K}.

Based on the available experimental data, several models with a small number of parameters have been developed to describe the astrophysical $S$-factors and derive the corresponding nuclear reaction rates. The standard and most widely used reference is CF88, which was originally implemented in the GENEC code and remains a common choice for stellar evolution modelling. For fusion reactions in particular, a phenomenological model deviating from CF88 at very low energies, the Hindrance (or HIN model), has recently been incorporated in GENEC. Both models are explored in this paper, and their implementation is described in detail in Sect.~\ref{subsubsection:data_driven_models}.
Alternatively, modern theoretical approaches can provide insight into the behaviour of the cross-sections at very low energies. These methods help constrain the rate trends or, in the absence of experimental data, offer predictive estimates that are used in stellar evolution codes (such as those available in the REACLIB database\footnote{https://reaclib.jinaweb.org/}).

For the three fusion reactions that are discussed in the present work (reactions \ref{eq:channels1}, \ref{eq:channels2}, and \ref{eq:channels3}), multiple frameworks have been used in the literature: the Time-Dependent Hartree-Fock method \citep[hereafter TDHF, e.g.][]{Bonche1978,Simenel2013,SIMENEL2018,Godbey2019,Scamps2019}, the Antisymmetrised Molecular Dynamics method \citep[][hereafter AMD]{TANIGUCHI2020,Taniguchi2021}, and the microscopic $\alpha$-cluster model \citep[e.g.][]{Depastas2023}.
However, these models are subject to limitations arising from the approximations and assumptions needed to perform the computations. For example, TDHF and microscopic $\alpha$-cluster models are unable to capture resonances in the cross-section and may omit relevant physical phenomena, especially for complex systems such as $\rm ^ {12}C+^{12}\!C$. 
A combination of direct measurements of nuclear reactions and model-based constraints on the energy dependence of the reaction rates is therefore necessary to advance towards a more accurate description of the nuclear reaction rates. The approach followed in this work is to compare the empirical (data-driven) models with the TDHF model, which is described in more detail in Sect.~\ref{subsubsection:tdhf}.

\subsubsection{Data-driven frameworks}
\label{subsubsection:data_driven_models}
We use as reference rates for the three fusion reactions the parametrised analytic formulae from CF88, a standard reference widely used in the literature.
In addition, we also employ the rates from the phenomenological hindrance scenario. This scenario accounts for the suppression in the fusion cross-sections at deep sub-barrier energies, a phenomenon observed in medium-mass fusion systems (hereafter simply referred to as \textit{systems}) \citep[e.g.][]{Jiang2007}. 
In Q<0 systems, the hindrance effect is clearly observed as a more rapid decline of the cross-section than expected, suggesting the existence of a maximum in the S-factor. 
Whether or not a similar suppression occurs in Q>0 systems  is still unclear. Although this trend has been experimentally investigated across a wide range of systems, it remains under debate whether such suppression also affects light fusion systems relevant to our study.

Assuming this interpretation is correct, \citet{Jiang2007} derived an analytical expression for the S-factor, fitted to experimental data across a broad range of systems. However, a complete physical explanation for the hindrance effect remains elusive. Adding a repulsive contribution to the ion-ion potential enables the reproduction of the observed behaviour. This repulsive term has been suggested to arise, at least partially, from a proper treatment of the Pauli repulsion \citep{Simenel2017,Umar2021}. 

Motivated by experimental constraints, we adopt the more realistic Hindrance+Resonance model proposed by M22 for the case of $\rm ^{12}C+^{12}\!C$. To describe the hindrance scenario for the $\rm ^{12}C+^{16}\!O$ and $\rm ^{16}O+^{16}\!O$ systems, we use the phenomenological formulation of the S-factor maximum provided by \citet{Jiang2007}. Starting from these expressions for the total S-factor, we derive the corresponding non-resonant nuclear reaction rates for each channel (proton, alpha and neutron).
The branching ratios for each channel are assumed to follow those of CF88, except for $\rm ^{12}C+^{12}\!C$, where experimentally derived branching ratios of 65-35~\% are adopted in favour of the $\alpha$ exit channel. We refer to this approach as the HIN(RES) scenario, which differs from the full Hindrance model (as explored in M22 and D24) by including the resonance only for the $\rm ^{12}C+^{12}\!C$ reaction given the lack of clear experimental resonance data within the Gamow window for $\rm ^{12}C+^{16}\!O$ and $\rm ^{16}O+^{16}\!O$. 
Figure~\ref{fig:rates_multi} shows the relative deviations from CF88 of the rates obtained under the HIN(RES) scenario as a function of temperature. 
The C- and O-burning temperature regimes are given for the 15 and 17~M$_{\odot}$ stellar models. 
For these masses, we expect the hindrance to impact the $\rm ^{12}C+^{16}\!O$ reaction while its influence on the $\rm ^{16}O+^{16}\!O$ reaction is expected to be minor, since the suppression in the cross-section predominantly occurs at temperatures below those relevant for core oxygen burning.  

\subsubsection{Time-dependent Hartree-Fock framework}
\label{subsubsection:tdhf}

As a complementary approach to the previous data-driven models we choose to explore a third scenario: the impact of nuclear reaction rates derived from the fully microscopic TDHF mean-field theory \citep[for a detailed description of this method see e.g.][]{SIMENEL2025}. TDHF simulations describe the time evolution of the occupied nucleon wave functions in the self-consistent mean field generated by the ensemble of nucleons. 
The many-body state is constrained to be an antisymmetrised product of single particle states (i.e. a Slater determinant) at all times. The main input is the energy density functional (EDF), which characterises the effective nucleon-nucleon interaction. Although most applications are based on the Skyrme EDF, recent calculations with the finite range Gogny interaction and including pairing correlations have also been performed by \citet{Scamps2019}. 

A significant advantage of the TDHF approach is that it describes both nuclear structure and dynamics on the same footing. In particular, EDF  parameters are only fitted to nuclear structure data and equation of state properties, without relying on input from nuclear reactions. Modern TDHF calculations can now be performed in full 3 dimensions, without imposing constraints on the systems shape evolution. The weakness of the real-time mean-field method is that it cannot account for the quantum tunnelling expected in the sub-barrier fusion. To overcome this, predictions are obtained by calculating the nucleus-nucleus potential at an energy just above the Coulomb barrier using the so-called density-constrained TDHF method \citep[DC-TDHF:][]{Umar2006}. 
The resulting potential is then used to compute barrier transmission probabilities at sub-barrier energies through one-barrier penetration methods. Importantly, DC-TDHF potentials account for crucial dynamical effects such as nucleon transfer \citep{godbey2017}, which may significantly impact sub-barrier fusion cross-sections. DC-TDHF calculations were performed in \citet{Godbey2019} for the case of $\rm ^{12}C+^{12}\!C$ system, which is also considered in the present work. Here, calculations based on the SLy4d \citep{kim1997}, UNEDF1 \citep{kortelainen2012}, and QMC \citep{stone2016} EDFs are used. The trends obtained are very similar for the three models and show an increase of the S-factor at low energies, in contrast to the Hindrance scenario (see App.~\ref{sec:c12c12}), though still several orders of magnitude lower than the predictions from THM discussed above. For our sensitivity study, we choose to adopt the predictions obtained with the Skyrme SLy4d functional. For the two other systems ($\rm ^{12}C+^{16}\!O$ and $\rm ^{16}O+^{16}\!O$), we perform analogous calculations using the same methodology as in \citet{Godbey2019}.

The resulting nuclear rates are shown in Fig.~\ref{fig:rates_multi} and compared to the CF88 and HIN(RES) rates. At the temperatures where core and shell C- and O-burning are predicted to occur, the TDHF rates are slightly higher than those of CF88 by less than 1~dex. In contrast, the Hindrance scenario yields significantly lower rates (up to 2 orders of magnitude at these temperatures) suggesting it may lead to the largest differences in the results.

\subsection{Neutron source nuclear reactions}
\label{sec:neutron-reactions}

\begin{table}[t]
    \centering
    \caption{Reference for updated reactions involved in the s-process nucleosynthesis.}
    \begin{tabular}{c|c|c}
    \hline \hline
    Reaction & D24 & This work \\
    \hline
    $\rm ^{13}C(\alpha,n)^{16}O$ & NACRE & \cite{Ciani2021} \\
    $\rm ^{14}N(\alpha,\gamma)^{18}F$ & NACRE & \cite{Iliadis2010} \\
    $\rm ^{17}O(\alpha,n)^{20}Ne$ & NACRE & FS22 \\
    $\rm ^{17}O(\alpha,\gamma)^{21}Ne$ & CF88 & FS22 \\
    $\rm ^{18}O(\alpha,\gamma)^{22}Ne$ & NACRE & \cite{Iliadis2010} \\
    $\rm ^{22}Ne(\alpha,n)^{25}Mg$ & \cite{Jaeger2001} & \cite{Adsley2021}$^\dagger$ \\
    $\rm ^{22}Ne(\alpha,\gamma)^{26}Mg$ & NACRE & \cite{Adsley2021}$^\dagger$ \\
    \hline
    \end{tabular}
    \label{tab:reactions_2}
    \tablefoot{NACRE: \cite{Angulo1999}. FS22: \cite{Frost-Schenk22}. $\dagger$ For $\rm T>1.33$~GK, we use the values from \cite{Longland2012} who extrapolated at high temperatures using the Hauser-Feshbach method. }
\end{table}

As the main contributor to free neutrons during massive star evolution, the rate of the $\rm ^{22}Ne(\alpha,n)^{25}Mg$ reaction has also been updated according to \citet{Adsley2021}. This reaction is in competition with the $\rm ^{22}Ne(\alpha,\gamma)^{26}Mg$, whose rate has also been updated based on the same reference. These updated rates differ significantly from previous evaluations \citep[][]{Longland2012,Wiescher2023}, and their reliability is still under debate, with additional experimental measurements expected to clarify the situation \citep[e.g.][]{2025JPhG...52b5201S}. For this study, we adopt these new rates to obtain a first estimation of its impact on massive star nucleosynthesis. Rates from \citet{Adsley2021} are approximately twice as high as those from \citet{Longland2012} and are expected to enhance the free neutron density during the C- to O-burning phases.

The second main contributor to free neutrons comes from the $\rm ^{13}C(\alpha,n)^{16}O$ reaction. In D24, we used the rates from the NACRE compilation \citep{Angulo1999}, but we now adopt the updated rates proposed by \citet{Ciani2021}. This new reference results in reduced neutron production while preserving a higher $\rm ^{13}C$ abundance throughout the He-burning phase. There is broad consensus regarding the reliability of these neutron source reaction rates.

From the original Netgen database and in line with recent references, we have updated four additional linked reactions, as listed in Tab.~\mbox{\ref{tab:reactions_2}}. Finally, we note that the update of the $\rm ^{12}C(\alpha,\gamma)^{16}O$ reaction should not alter significantly the s-process nucleosynthesis, as demonstrated by \citet{deBoer2017}.

\section{Nuclear reaction rates impact: From C-burning to O-burning}
\label{sect:class}

\subsection{General evolution}
\begin{figure}[t]
    \center
    \includegraphics[width=90mm]{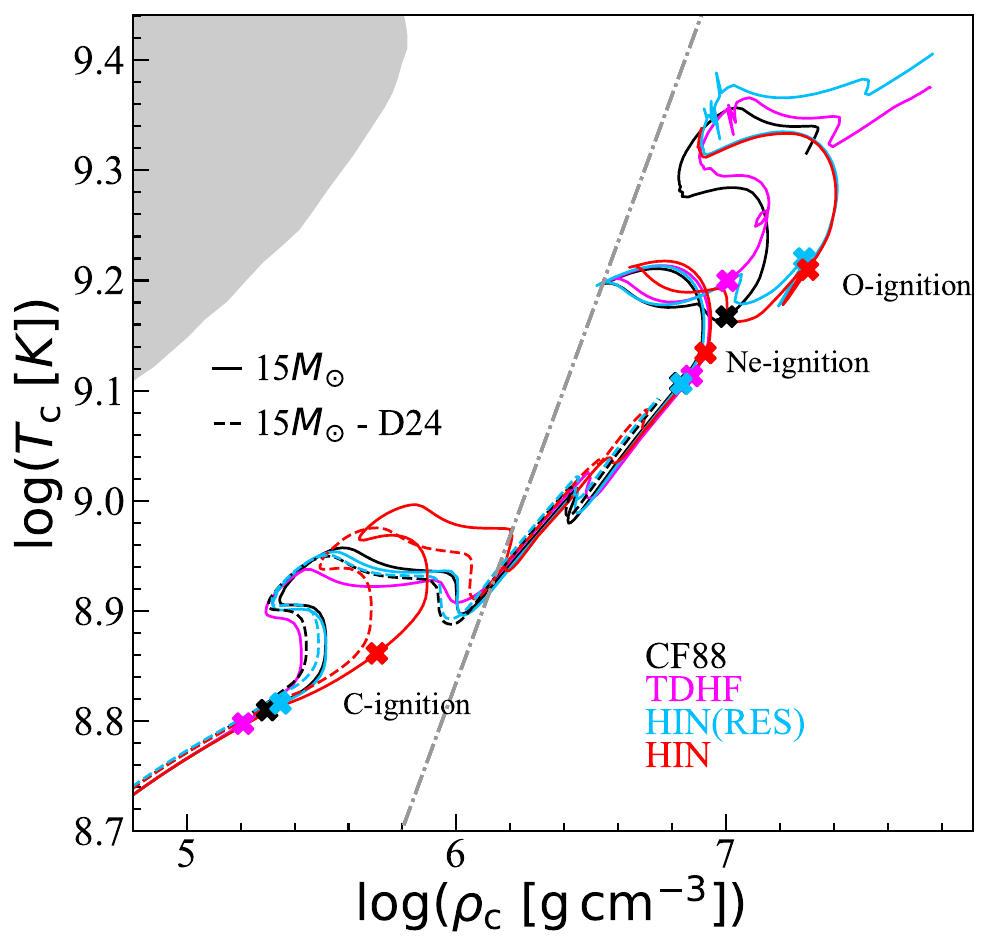}
    \caption{Evolution of central temperature ($\rm T_C$) vs. central density ($\rm \rho_C$) for rotating models of 15~M$_{\odot}$ at solar metallicity, shown for the three nuclear reaction rate references considered in this work along with the full Hindrance scenario (HIN model) with each model colour-coded accordingly. Nuclear ignition points are marked by crosses. Models from D24 are indicated in dashed-lines for the CF88, HIN, and HIN(RES) models. 
    The dash-dotted grey line indicates the boundary for degenerate conditions inside the core, while the shaded grey area denotes the domain affected by electron-positron pair instability. 
    }
    \label{fig:rhoctc15}
 \end{figure}

In this section, we compare the evolutionary outcomes of the rotating models under each of the three nuclear reaction rate scenarios, CF88, HIN(RES) and TDHF, described in detail in Sect.~\ref{sect:nucl}. Figure~\ref{fig:rhoctc15} shows the evolution of the central temperature and density of our 15~M$_{\odot}$ models for each of the three nuclear reaction rates references. Models from D24 are also included in the figure for comparison. Compared to the D24 models (computed up to Ne-ignition), the new models exhibit a hotter and denser core. This is a consequence of the updated reaction rates for $\rm ^{12}C(\alpha,\gamma)^{16}\!O$ during the He-burning phase and the revision of the EOS in the advanced phases. As observed in D24, the C-ignition conditions depend on the reaction rates. In the new models, ignition occurs at lower temperatures in the TDHF and at higher temperatures in the HIN(RES), relative to the CF88 model. We note that the overall shift to higher temperatures in the new models, compared to D24 reduces the difference between the CF88 and HIN(RES) models, as the C-burning temperatures of the 15~M$_{\odot}$ star are now closer to the peak of the resonance, where the CF88 and HIN(RES) rates converge (see the top right panel in Fig.~\ref{fig:rates_multi}). A similar behaviour can be observed for O-ignition where the two models following the hindrance scenario show a shift towards ignition in a denser and hotter core. 

\begin{figure*}[t]
    \includegraphics[width=90mm]{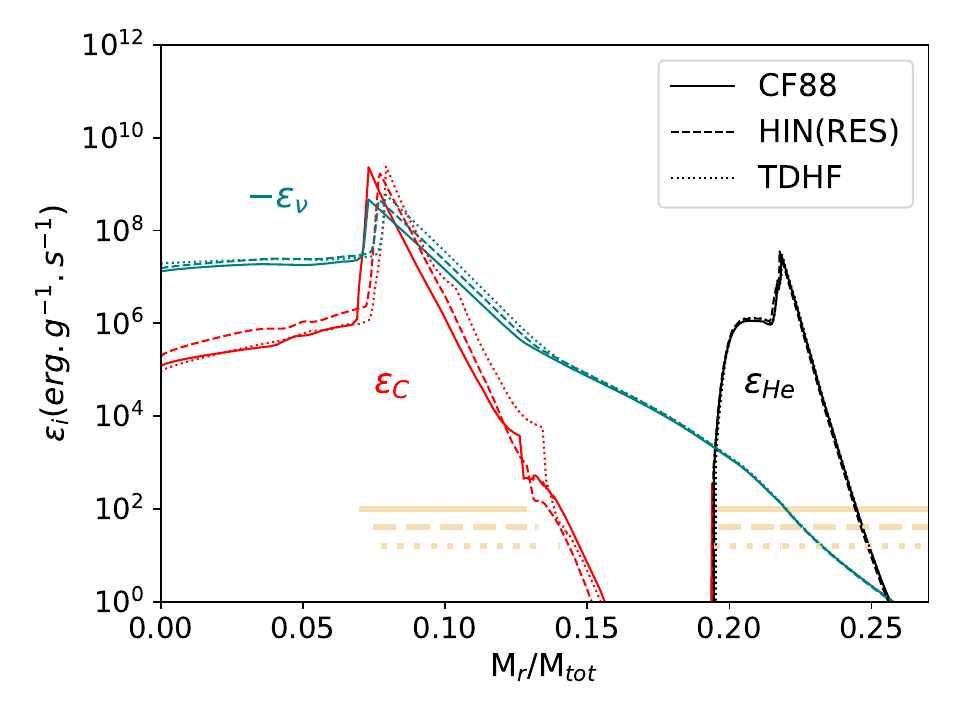}
    \includegraphics[width=90mm]{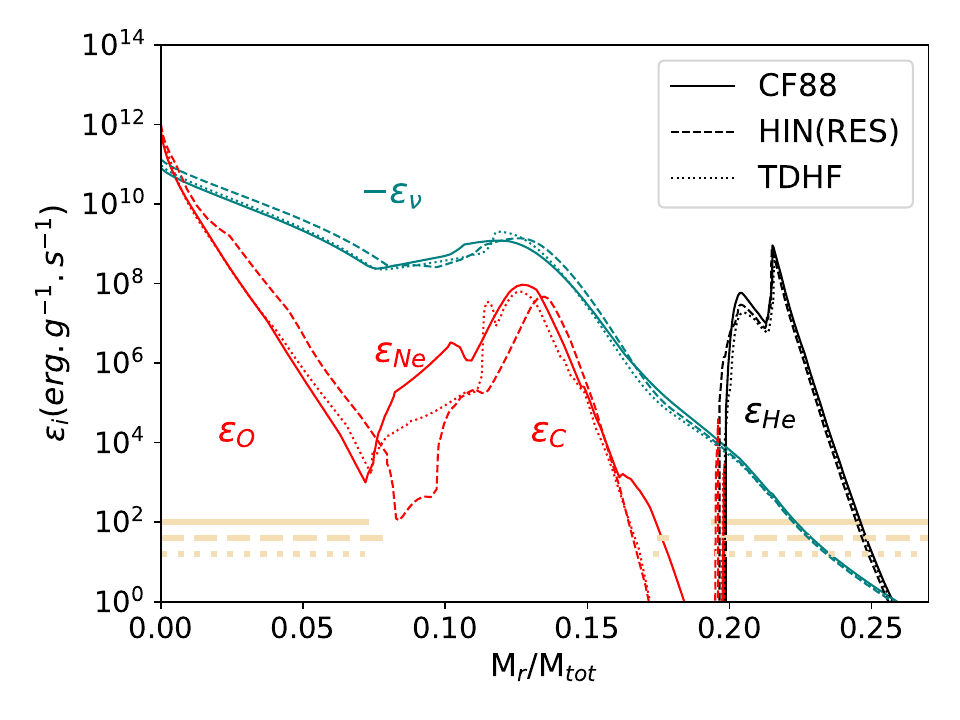}
    \caption{Energy production rates $\epsilon$  [erg.$\rm g^{-1}$.$\rm s^{-1}$] inside a 15~M$_{\odot}$ rotating star for varying reaction rates. Left: Energy during shell C-burning, approximatively 3000 years after core C-ignition. Right: Energy during core O-burning at central $X_{^{16}O} = 0.3$. The pale areas in orange indicate where the gravitational energy is negative, \textit{i.e.} corresponding to an expansion.
    }
    \label{fig:energy_profiles}
 \end{figure*}

Changes in the rates can also be observed from the perspective of the energy output associated with the various reactions during the evolution. In Fig.~\ref{fig:energy_profiles}, we show the total energy production profile for a rotating 15~M$_{\odot}$ star during the shell C-burning and core O-burning phases, for each of the nuclear rate prescriptions explored in this study. Nuclear energy produced in the core or in shell is compared to the neutrino energy loss, which becomes significant during the core C-burning phase. Moderate differences between models are observed, mainly reflecting the readjustment of the stellar structure which compensates the lower HIN(RES) rates by higher central temperatures and shorter burning lifetimes, resulting in a modest increase in central energy production.

\subsection{Burning lifetimes}

\begin{table}[t!]
    \centering
    \caption{Core-burning lifetimes (in years) for rotating and non-rotating stars of 15, 17, and 20~M$_{\odot}$ as a function of the adopted nuclear reaction rates. 
    }
    \begin{tabular}{c|c|c|c|c}
        \multicolumn{5}{c}{C-burning} \\
         \hline \hline
         Model & $\rm V_{ini}/V_{crit}$ & CF88 & HIN(RES) & TDHF \\
         \hline
         15~M$_{\odot}$ & 0 & 4024 & 4014 & 5426 \\ 
            & 0.4 & 3014 & 2809 & 3759 \\ 
         17~M$_{\odot}$ & 0 & 2360 & 2372 & 3092 \\ 
          & 0.4 & 1873 & 1651 & 2134 \\ 
         20~M$_{\odot}$ & 0 & 1059 & 1042 & 1418  \\ 
          & 0.4 & 795 & 830 & 1070 \\ 
         \hline
    \end{tabular} \\
    \begin{tabular}{c|c|c|c|c}
        \multicolumn{5}{c}{Ne-burning} \\
         \hline \hline
         Model & $\rm V_{ini}/V_{crit}$ & CF88 & HIN(RES) & TDHF \\
         \hline
         15~M$_{\odot}$ & 0 & 3.71 & 2.78 & 2.26 \\
            & 0.4 & 2.26 & 2.41 & 1.51 \\
         17~M$_{\odot}$ & 0 & 3.2 & 5.82 & 2.75 \\
          & 0.4 & 0.52 & 0.49 & 0.50 \\
         20~M$_{\odot}$ & 0 & 0.65 & 0.94  & 0.25 \\
            & 0.4 & 0.75 & 0.32 & 1.30  \\
         \hline 
    \end{tabular} \\
    \begin{tabular}{c|c|c|c|c}
        \multicolumn{5}{c}{O-burning} \\
         \hline \hline
         Model & $\rm V_{ini}/V_{crit}$ & CF88 & HIN(RES) & TDHF \\
         \hline
         15~M$_{\odot}$ & 0 & 1.78 & 0.68 & 1.16 \\
            & 0.4 & 1.04 & 0.59 & 1.20 \\
         17~M$_{\odot}$ & 0 & 0.80 & 0.70 & 0.73 \\
          & 0.4 & 0.52 & 0.42 & 0.63 \\
         20~M$_{\odot}$ & 0 & 0.49 & 0.26 & 0.55 \\
            & 0.4 & 0.45 & 0.18 & 0.37  \\
         \hline 
    \end{tabular}
    \label{tab:lifetime}
\end{table}

\begin{figure}[t]
    \center
    \includegraphics[width=90mm]{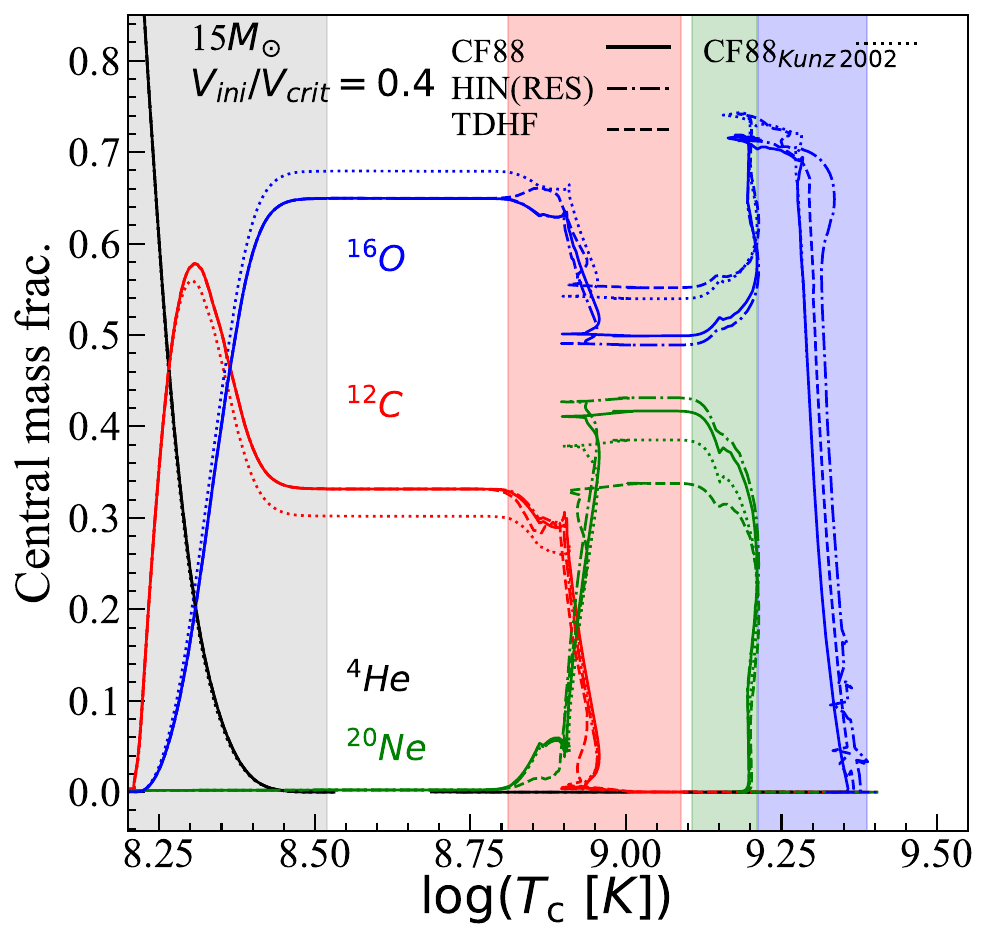}
    \caption{Mass fraction of central abundances for a rotating 15~$M_{\odot}$ from He-burning to O-burning, shown for the three nuclear models: CF88, HIN(RES), and TDHF (line-coded). For comparison, a CF88 model using the rates from \citet{Kunz2002} is shown in dotted lines. Shaded areas show the successive nuclear burning phases with colours corresponding to He-, C-, Ne-, and O-burning (grey, red, green, and blue, respectively).
    }
    \label{fig:evolelem}
 \end{figure}

The durations of the various nuclear burning phases are especially sensitive to changes in reaction rates. We observe significant differences in the present work compared to the values reported in D24, which we attribute to the updated reference for the $\rm ^{12}C(\alpha,\gamma)^{16}\!O$ reaction. As discussed in Sect.~\ref{sect:cago}, the rates from \citet{deBoer2017} increase the ratio of $\rm ^{12}C/^{16}O$ from about 0.45 to about 0.51, in the case of 15~M$_{\odot}$ model, as can be seen in Fig.~\ref{fig:evolelem}. In parallel with this increase, the change in $\rm ^{12}C(\alpha,\gamma)^{16}\!O$ rates also leads to a reduction in the amount of $\rm ^{20}Ne$ produced at He-exhaustion. Overall, the transition from the \citet{Kunz2002} to the \citet{deBoer2017} rate results in shorter He-, Ne-, and O-burning lifetimes, and an extended C-burning phase.

Regarding the impact of fusion reaction rates variations under the three scenarios, Tab.~\ref{tab:lifetime} presents the lifetimes for the C-burning, Ne-burning, and O-burning phases for rotating and non-rotating models across the explored mass-range. 
To determine these lifetimes we define the ignition point as the moment when the central mass fraction of the main fuel element decreases by 3$\times 10^{-3}$ from its maximum value $\rm X_{fuel,max}$. 
We define the point of exhaustion as the time when the central mass fraction reaches minimum $\rm X_{fuel,min}$ set to $10^{-5}$ for C, and $10^{-3}$ for Ne and O. In other words, the phase lifetime is: $\rm t(X_{fuel,max}-3\times 10^{-3}) - t(X_{fuel,min})$. The lower rates from the HIN(RES) result in comparable or slightly shorter C-burning lifetimes (with a numerical uncertainty of $\approx 100$ years). In contrast, the higher reaction rates from the TDHF model lead to a significant increase in C-burning lifetime for all models relative to the CF88 reference. This result may appear counter-intuitive, as one might expect lower rates to yield longer lifetimes due to less efficient burning. However, as observed by M22 and D24, reduced fusion probabilities decreases the energy output in the burning core. To maintain hydrostatic equilibrium, the star then responds by contracting its core, which in turn heats up (see the crosses in Fig.~\ref{fig:rhoctc15}), thus boosting the burning and reducing the phase lifetime. Additionally, the lower fusion probability delays ignition to higher densities, further shortening lifetime.

For the Ne-burning phase, differences between models are less significant, and no clear monotonic trend emerges in response to the changes in $\rm ^{12}C+^{12}\!C$ reaction rates, as previously noted by \citet{Bennett2012}. Indeed, for the advanced Ne- and O-burning phases, the lifetime is governed by a more complex interplay between the central temperature reached at the end of the C-burning and the residual abundances of Ne and O.

The O-burning lifetimes are notably shorter in the HIN(RES) models, due to both a lower amount of available $^{16}$O and the correspondingly reduced $\rm ^{16}O+^{16}\!O$ reaction rates. This result reflects the same underlying mechanism discussed for the $\rm ^{12}C+^{12}\!C$ reaction, with  ignition conditions shifted to higher temperatures and densities, as shown in Fig.~\ref{fig:rhoctc15}. As suggested by Fig.~\ref{fig:rates_multi}, the CF88 and TDHF models yield similar O-burning lifetimes due to their nearly identical rates at O-burning temperatures.

Finally, we note that both rotating and non-rotating models exhibit trends consistent with those reported in D24; we refer the reader there for further details.

\subsection{Stellar structure}

\begin{figure*}[t]
    \includegraphics[width=90mm]{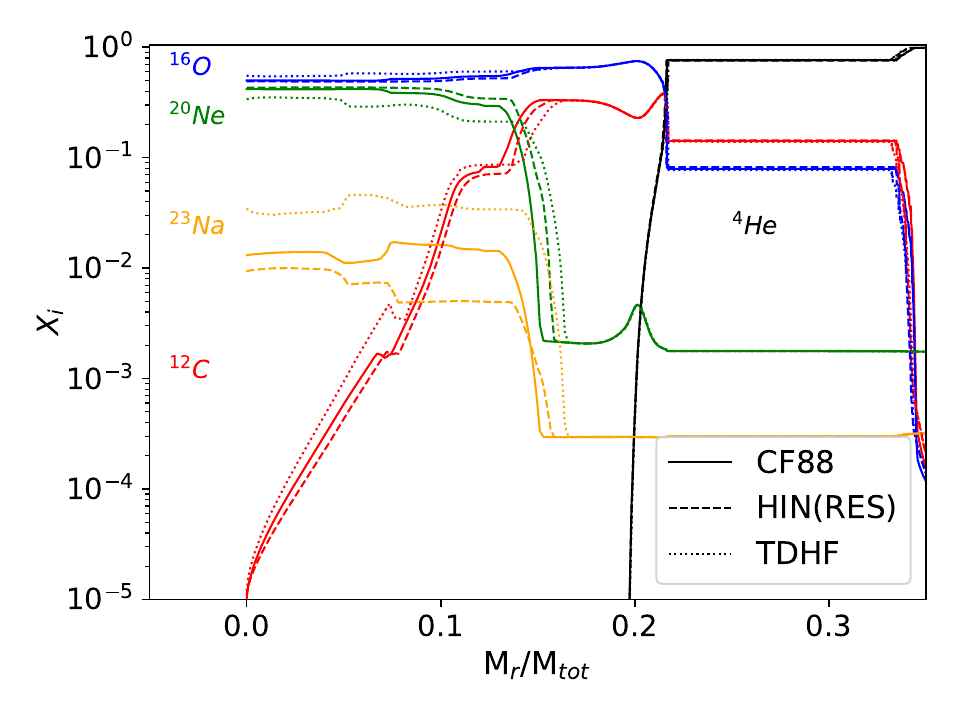}
    \includegraphics[width=90mm]{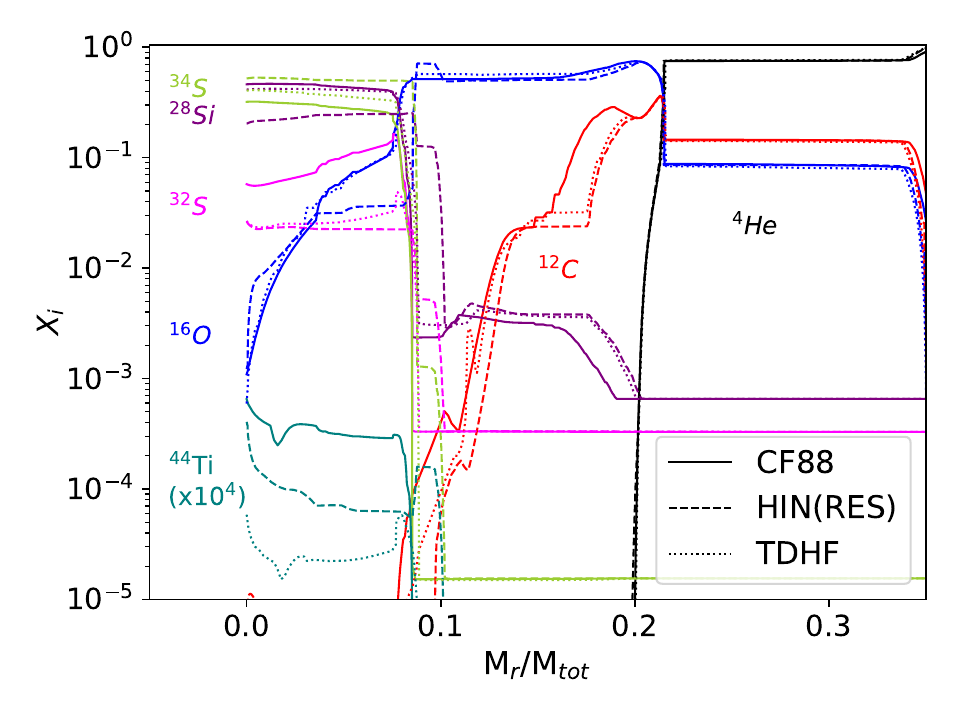}
    \caption{Profiles of abundances of $^4$He, $^{12}$C, $^{16}$O, $^{20}$Ne, $^{23}$Na, $^{28}$Si, $^{32}$S, $^{34}$S, and $^{44}$Ti (colour-coded) for a rotating 15~M$_{\odot}$, shown for different reaction rates scenarios distinguished by line style. Left panel: End of the C-burning phase. Right panel: End of the O-burning phase.
    }
    \label{fig:abundance_profiles_rates}
 \end{figure*}

The chemical structure of the most abundant elements is shown in Fig.~\ref{fig:abundance_profiles_rates} at the end of the C-burning and O-burning phases for a 15~M$_{\odot}$ star under each of the nuclear reaction rate scenarios considered. At the end of the C-burning phase, all models predict similar CO core sizes of about 2.50~M$_\odot$. In the centre, chemical abundances are impacted by the change in reaction rates. The most striking difference is observed for $\rm ^{23}Na$, whose abundance is three times higher in the TDHF model than in the other two scenarios. This enhancement is attributed to a lower efficiency in the conversion of $\rm ^{23}Na$ into $\rm ^{20}Ne$ via subsequent (p, $\alpha$) reactions, due to the lower central temperature.
The extent of the carbon shells is also significantly impacted by the choice of reaction rates, as already noted in M22. In App.~\ref{fig:kippen}, we show the Kippenhahn diagrams for the 20 $M_{\odot}$ model under each scenario. It is clear that the highest rates from TDHF produce the least extended convective C-burning shell while the lowest rates from HIN(RES) result in a more extended shell. 

The right panel of Fig.~\ref{fig:abundance_profiles_rates} shows abundance profiles at the end of the O-burning phase. $\rm ^{28}Si$, $\rm ^{32}S$, and $\rm ^{34}S$ are the main products of this phase, formed through secondary reactions involving the primary $\rm ^{28}Si$ and $\rm ^{31}P$ products mentioned in Eq.~\ref{eq:channels3}, leading notably to a large abundance of $\rm ^{34}$S, consistent with finding by \citet{Farmer2016}. This amount of $^{34}$S is lower in the CF88 model compared to the TDHF and HIN(RES) cases. Moreover, the lower HIN(RES) rates result in a reduced central abundance of $\rm ^{28}Si$ relative to CF88 and TDHF. The resulting Si-core mass, M$\rm _{Si}$, is smaller for the CF88 and TDHF models ($\rm \approx 1.10 M_{\odot}$) than for HIN(RES) ($\rm \approx 1.20 \, M_{\odot}$, defined by $\rm X_{Si}+X_{S}$ < 0.2), in agreement with the results of \citet{2004A&A...425..649H} at the end of O-burning. Other elements are also affected, such as $^{44}$Ti, which is especially less abundant in the TDHF model compared to the other two scenarios, with mass fraction reaching only $\sim 6\times10^{-9}$, instead of 4-7$\times10^{-8}$ in the CF88 and HIN(RES) models.

To sum up, $\rm ^{12}C+^{12}\!C$ and $\rm ^{16}O+^{16}\!O$ rate variations impact the chemical structure, central abundances of several elements, and the Si core mass at the end of the O-burning phase. However, they have a negligible influence on the size of the CO core at this evolutionary phase.

\section{Nucleosynthesis}
\label{sect:nucleo}

Neutron capture processes are believed to be at the origin of most heavy elements in the Universe \citep[see e.g. the review by][]{Arcones2023}. One of these, the slow neutron-capture process (s-process), develops in massive stars, at neutron densities of the order of $\rm N_n = 10^5-10^{10}$ $\rm cm^{-3}$. It occurs mainly during the core He-burning phase \citep[e.g.][]{Langer1989, Prantzos1990, Raiteri1991a} with $\rm ^{22}Ne(\alpha,n)^{25}Mg$ as the main neutron source. The $\rm ^{13}C(\alpha,n)^{16}O$ reaction produces a short neutron burst at the onset of core He-burning, barely producing heavy elements. These s-process elements are subsequently transported to the outer stellar layers over the relatively long duration of the He-burning phase. During the brief advanced evolutionary stages, the C-burning shell also contributes to the s-process \citep[e.g.][]{Raiteri1991b, The2007, Pignatari2013}. Minor contributions are also expected during the core C-burning phase \citep[e.g.][]{Bennett2012,Pignatari2013}. In this specific phase, rapid transport of nucleosynthetic products from the core to the outer shells should be involved to enable this process to contribute to the chemical galactic enrichment. During the core O-burning phase, trans-iron elements are destroyed via photodisintegration reactions. In addition, rotation is is known to boost the s-process during core He-burning by promoting the synthesis of primary $^{13}$C and $^{22}$Ne \citep{Pignatari2008, Frischknecht2016,Choplin2018, Limongi2018, Banerjee2019}. The impact of variations in nuclear reaction rates and rotation parameters manifests mainly through changes in central temperatures throughout stellar evolution and modifications in the burning lifetimes, both of which influence the neutron exposure. Indeed, the two main neutron sources for s-process in massive stars--$^{13}$C and $^{22}$Ne--contribute in different proportions depending on the temperature and evolutionary stage. In this context, we have also updated seven important reactions for s-process nucleosynthesis, as listed in Tab.~\ref{tab:reactions_2} and introduced in Sect.~\ref{sect:nucl}.

\subsection{One-zone code for nucleosynthesis}

We explore the implications for nucleosynthesis using the one-zone code described in \citet{Choplin2016}, and previously applied in \cite{Taggart2019,Williams2022,Frost-Schenk22}, M22 and D24. The nuclear network has been extended to track 1454 isotopes (against 737 in the previous version in D24) in order to capture photodisintegration reactions during the core O-burning phase.
In this approach, only a single thermodynamic trajectory in temperature and density is followed (either core or shell), and transport processes, such as convection, are neglected. This simplified method provides reliable predictions for the s-process nucleosynthesis during the core He-burning phase \citep[e.g.][]{Frost-Schenk22}. During these advanced evolutionary phases, rotational transport is still active but exerts a weaker mixing influence due to the short timescales in late stellar evolution \citep[as shown, e.g. in Fig.~2 by ][]{Choplin2017}. Thus, this approach is sufficient sufficient to capture and interpret the relative effects introduced by variations in nuclear reaction rates. Numerous studies have investigated s-process nucleosynthesis in the core and the shell regions during the advanced phases of evolution \citep{Arcoragi1991,2007PhRvC..76c5802G,The2007,2017MNRAS.469.1752N}.

\subsection{s-process nucleosynthesis}

\begin{table}[t]
    \centering
    \caption{Peak neutron density $\rm \rho_n^{max}$ ($\rm cm^{-3}$) and neutron exposure $\tau_n$ (mbarn$^{-1}$) during the first and second C-burning shell episodes for each of the three nuclear references: CF88, HIN(RES), and TDHF. 
    }
    \begin{tabular}{l|c c | c c}
        Model & $\rm \rho_n^{max, 1}$ & $\rm \rho_n^{max, 2}$ & $\rm \tau_n^{1}$ & $\rm \tau_n^{2}$ \\
        \hline
        CF88 & 1.4$\times 10^{9}$ & 3.1$\times 10^{10}$ & 7.4$\times 10^{-2}$ & 4.5$\times 10^{-2}$\\
        HIN(RES) & 9.8$\times 10^{8}$ & 1.3$\times 10^{10}$ & 8.9$\times 10^{-2}$ & 3.0$\times 10^{-2}$\\
        TDHF & 3.6$\times 10^{9}$ & 5.3$\times 10^{11}$ & 8.1$\times 10^{-2}$ & 8.3$\times 10^{-2}$
    \end{tabular}
    \tablefoot{Values are computed following the same method as in D24.}
    \label{tab:nucleo}
\end{table}

\begin{figure}[t]
    \center
    \includegraphics [width=100mm,trim=10mm 00mm 00mm 15mm, clip]{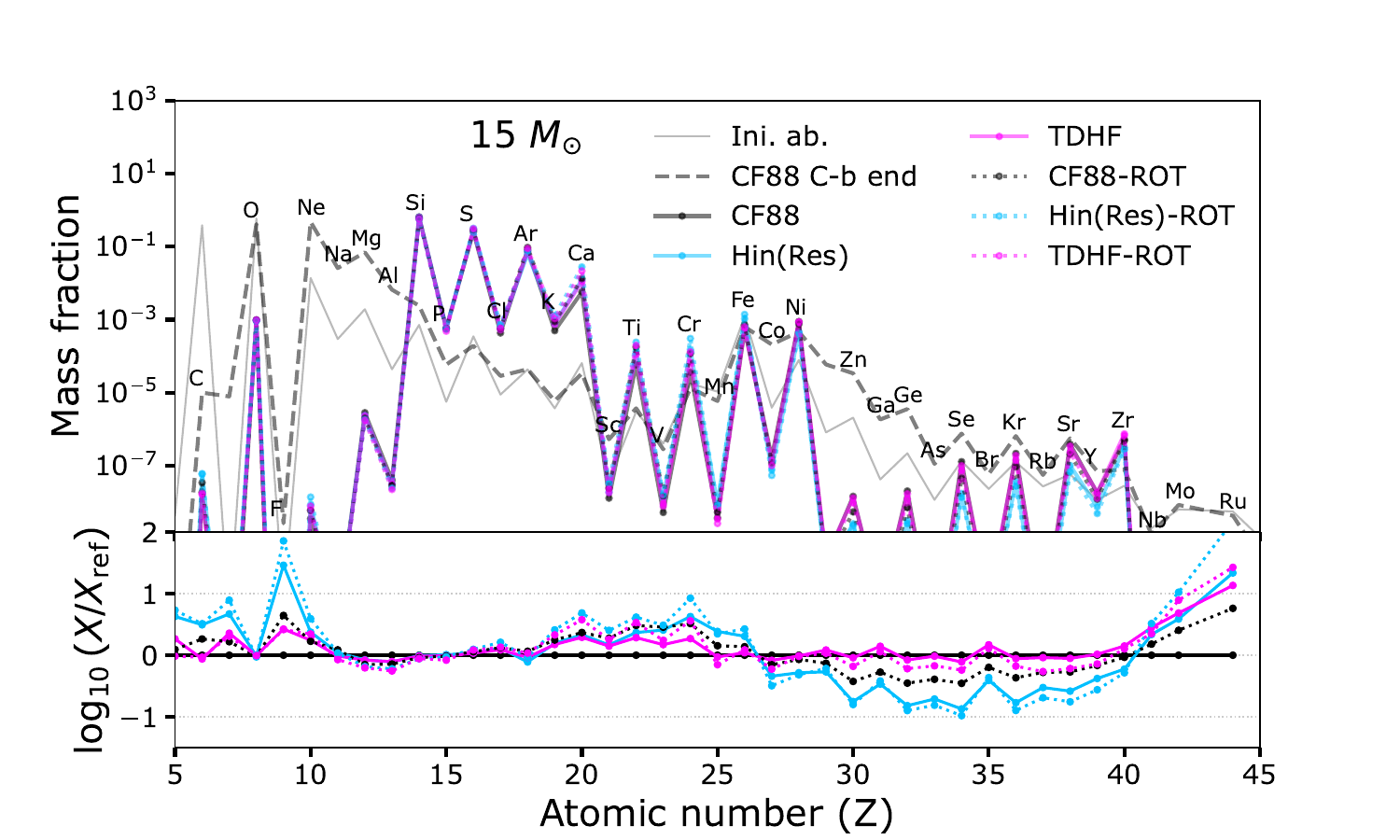}
    \caption{Mass fractions at the end of the core O-burning from one-zone calculations based on central conditions of a 15~M$_{\odot}$ star with and without rotation, shown for three nuclear rate references (colour-coded). The dashed grey line shows the abundances at the end of the core C-burning from the one-zone calculation of the non-rotating CF88 model, while the solid grey line indicates the initial composition at the onset of core C-burning. Lower panels display the abundances normalised to those of the non-rotating CF88 case.
    }
    \label{fig:nucleo_Ob}
 \end{figure}
 
We now explore the isotopic abundance predictions obtained from the one-zone code at the end of core O-burning and shell C-burning in our 15\,M$_{\odot}$ rotating models, where the slightly lower temperatures reached are expected to enhance the sensitivity to differences among the nuclear reaction rates prescriptions (see Fig.\,\ref{fig:rates_multi} or D24). We first examine the chemical evolution of the core, following the same approach as in M22 and D24, by tracking the central density and temperature throughout the successive core C-, Ne- and O-burning phases. Figure~\ref{fig:nucleo_Ob} shows the production of both direct C-burning products, such as Ne, Na, and Mg, and trans-iron elements during the core C-burning phase. 
After Ne-burning, the subsequent O-burning phase leads to the destruction of most trans-iron elements, and the synthesis of typical O-burning products, including Si, S, Ar, Ca, and Ti, as also observed in the right panel of Fig.~\ref{fig:abundance_profiles_rates}. 
At the end of the core O-burning phase, heavy elements exhibit either low ($30<Z<40$) or almost null ($Z>40$) abundances (coloured lines in Fig.~\ref{fig:nucleo_Ob}). This is due to photo-dissociation reactions such as $(\gamma,n)$, $(\gamma,p)$, and $(\gamma,\alpha)$, which become increasingly efficient at $T \gtrsim 1.8$~GK \citep[e.g.][]{arnould03}, typically reached during the O-burning phase. The sawtooth structure reflects odd-even effects in the nuclear binding energies due to pairing, and is clearly visible in the abundance pattern at the end of O-burning.
Beyond the low abundance of fluorine ($Z=9$), the largest deviations -- up to $\sim$\,1\,dex -- relative to the CF88 reference are obtained for the HIN(RES). These feature a modest overproduction of light elements between Ca ($Z=20$) and Fe ($Z=26$), and an underproduction of s-process elements at $Z > 30$, such as Ge, Se, or Sr. We mainly observe the consequences of the lower rates associated with the Hindrance scenario for the $\rm ^{16}O+^{16}\!O$ reaction, which leads to a shorter burning phase and reduced neutron exposure. This mirrors the behaviour seen in D24 for the $\rm ^{12}C+^{12}\!C$ reaction during core C-burning phase. It is worth noting that the cut-off for $Z>40$ is unaffected by the choice of reaction rates.

In all models, the s-process distribution in the core is further modified during carbon-burning shell episodes. To explore the associated nucleosynthesis, we have extracted the evolution of density and temperature from the first two C-burning shells, which occur between the end of core C-burning and onset of core Ne-ignition (see technical details in App.~\ref{sec:nucleosynthesis-trajectories}). Compared to the helium core, the carbon shells overproduce specific isotopes due to the higher neutron densities, reaching up to $10^{10}$cm$^{-3}$ within the second C-burning shell, as shown in Tab.\,\ref{tab:nucleo}. For instance, in the Sm, Eu, and Gd region, we observe an overproduction of $\rm ^{152,154}Gd$, corresponding to the opening of branching points at A\,=\,151 and A\,=\,154. This enhancement is accompanied by a reduction in the abundance of intermediate isotopes along the same nucleosynthetic path (e.g. $\rm ^{151,153}Eu$ is depleted by an order of magnitude). 
However, differences between models remain modest. For example, the $\rm ^{152}Gd$ isotope is overproduced by a factor of 10 in the HIN(RES) scenario, compared to a factor of 2 in the TDHF scenario, relative to its abundance at the end of the He-burning phase.
The reasons for the small changes observed are twofold. First, the corresponding temperature and density trajectories do not differ significantly enough across the three explored cases to drive substantial nucleosynthetic differences (see App.~\ref{fig:shell_burning_trajectories}). 
Second, and more generally, for all our 15~M$_{\odot}$ rotating models, the s-process is not very efficient, as quantitatively demonstrated by the peak neutron densities and the total neutron exposures given in Tab.~\ref{tab:nucleo}, with maximum exposures reaching only $\sim$0.1\,mbarn$^{-1}$ during the first C-burning shell. 

\subsection{Variation of the $\rm ^{12}C+^{16}\!O$ reaction}

Finally, this paper is also intended to guide further experimental determinations of reaction cross-sections. To this end, we performed a sensitivity study for the $\rm ^{12}C+^{16}\!O$ reaction. Unlike the main fusion reactions (\ref{eq:channels1} and \ref{eq:channels3}), this reaction does not contribute significantly to energy generation. As such, it is only relevant primarily at the end of core C-burning or during shell C-burning, once the required thermodynamic conditions, low $\rm ^{12}C$ abundance, high $\rm ^{16}O$ abundance, and suitable temperature are met \citep{Iliadis2010}. We investigate nucleosynthesis in the first C-burning shell of our 15~M$_{\odot}$ model, where the differences among reaction rates are most pronounced. We conduct the sensitivity study by arbitrarily multiplying or dividing the CF88 reaction rates for $\rm ^{12}C+^{16}\!O$ by constant factors.
We ind that only an extreme rate variation (by a factor $\sim\,10^{4}$ would significantly affect the nucleosynthesis. \\ Specifically, increasing the rate by this factor results in an overproduction of Al and Si by factors of 2.40 and 1.25, respectively, while decreasing the rate yields underproduction of Mg, Al, and Si by $\sim$20. We conclude that the differences in $\rm ^{12}C+^{16}\!O$ reaction rates among the CF88, HIN(RES), and TDHF scenarios (see Fig.\,\ref{fig:rates_multi}) are too small to significantly affect nucleosynthesis. Given the current measurements of nuclear reaction rates and predictions for this reaction, no strong impact is expected from improved accuracy, unless previously unidentified resonance enhancing the rate by several orders of magnitude is discovered at astrophysical energies.

\section{Remnant predictions}
\label{sect:remnant}

A major goal of massive star modelling is to determine which stars explode and which collapse into black holes \citep[e.g.][]{Chieffi2020,Chieffi2021}. It is essential for constraining black hole (BH), neutron star (NS), and supernova (SN) populations synthesis models. Recent works, such as \citet{Ugolini2025} investigate the link between progenitor mass and the remnant mass produced by core collapse supernovae.

Several attempts to estimate the type of remnant (and the associated explosion type) resulting from massive star evolution have been proposed in the literature \citep[e.g.][]{Maltsev2025}. Some approaches require knowledge of quantities near the onset of core collapse \citep[e.g.][]{Ugliano_2012,Ertl_2016,Muller_2016}, while others infer the remnant type based on the CO core properties {\citep[e.g.][]{Patton2020,Maltsev2025}. Since our models are evolved up to the end of the O-burning phase, we estimate their fate based on properties at this state.
The work by \citet{Patton2020} sought to systematically predict whether the remnant would be a BH or NS, and to estimate its mass, using only the CO core mass and central carbon mass fraction at He-exhaustion as diagnostic parameters in non-rotating stars. From these two parameters, they build a remnant classification map of explosion predictions. While such a two-parameter model cannot fully capture the complex dynamics of core collapse, and state-of-the-art multi-dimensional simulations would be required for a detailed outcome \citep[see e.g.][]{Wang_2022MNRAS.517..543, Burrows_2024}, it is still possible to use these parameters to explore how variations in the input physics between models could affect the final fate of massive stars.

\subsection{Updated $\rm ^{12}C(\alpha,\gamma)^{16}O$ reaction rates}

In Sect.~\ref{sect:nucl} we have shown that updating the reaction rates of $\rm ^{12}C(\alpha,\gamma)^{16}O$ from those of \citet{Kunz2002} to those of \citet{deBoer2017} changes the amount of $^{12}$C left in the core at He-exhaustion, while leaving the CO core mass nearly unchanged (see App.~\ref{sec:remmnant_properties} for the exact values of $\rm M_{CO}$ and $\rm X_{^{12}C}$). Using the remnant classification of \citet{Patton2020} based on the \cite{Kunz2002} rates, we find that all models would lead to NS formation, with the exception of the 20\,M$_{\odot}$ model, which collapses into a BH. 
In Fig.~\ref{fig:all_MCO}, we overlay the position $\rm (M_{CO},X_{^{12}C})$ of our models onto the remnant-type map from \cite{Patton2020}. The general trend resulting from adopting the \citet{deBoer2017} rates is clear: they yield a higher central carbon mass fraction $\rm X_{^{12}C}$ and a slightly smaller CO core mass. Additionally, as the initial mass increases, $\rm M_{CO}$ increases while $\rm X_{^{12}C}$ decreases, owing to He-burning occurring at higher temperatures. From the remnant map of \citet{Patton2020}, we observe that, broadly speaking, a higher $\rm M_{CO}$ or a lower $\rm X_{^{12}C}$ correlates with an increased likelihood of BH formation. However, the map contains multiple islands of explosiveness as seen in the background of Fig.~\ref{fig:all_MCO}. The variation in the central value of $\rm X_{^{12}C}$ resulting from the updated reaction rates is large enough to change the predicted remnant type according to \citet{Patton2020}. In contrast, using the prescription of \citet{Maltsev2025}, all our models would be classified as NS progenitors, since all their CO core masses are below 6.6\,M$_{\odot}$, the threshold below which they predict NS formation for all models. 

\begin{figure}[t]
    \center
    \includegraphics[width=90mm]{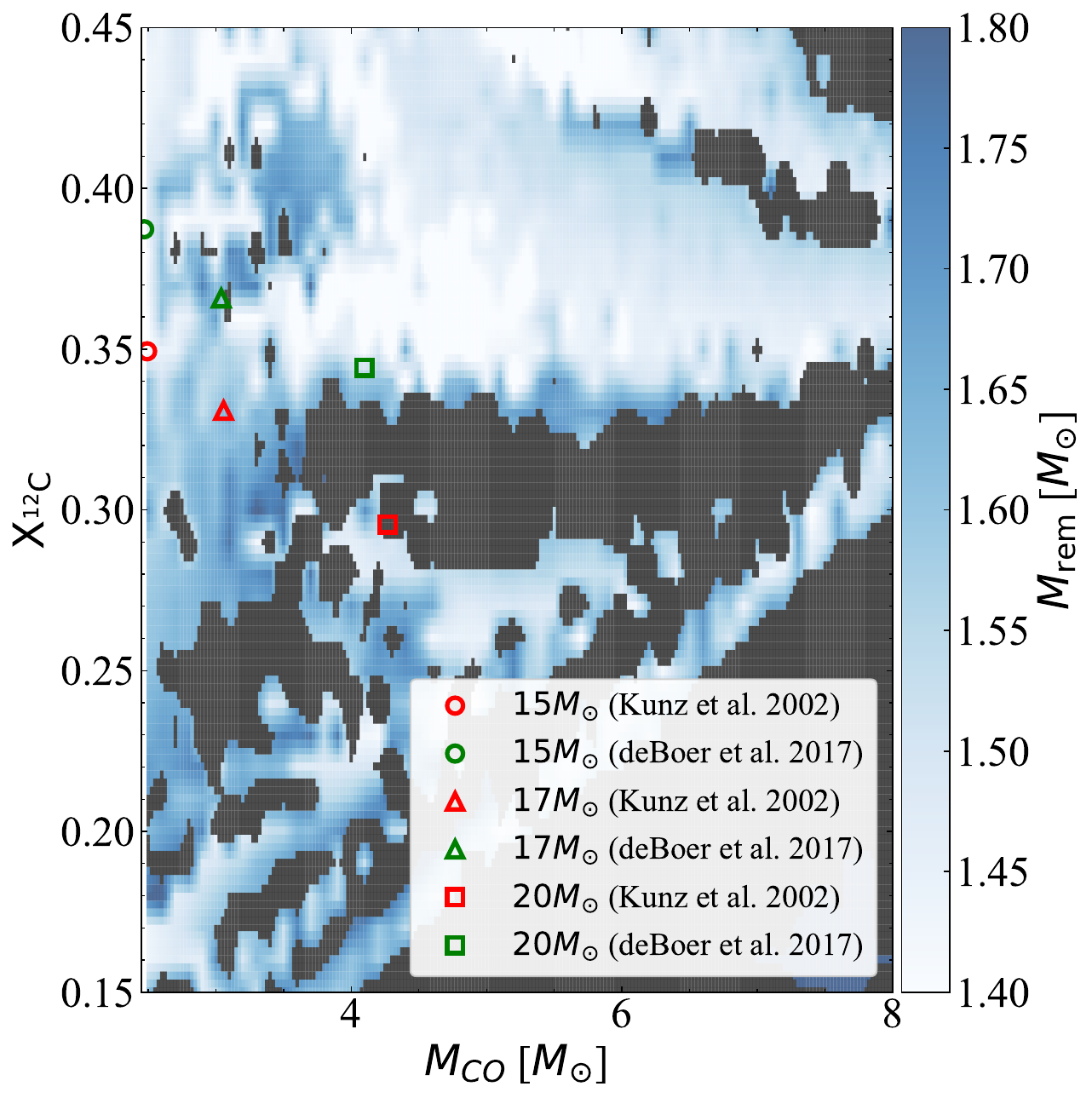}
    \caption{
    Position of our models in the $\rm(M_{CO},X_{^{12}C})$ parameter space for all initial masses using the rates from \citet[][red]{Kunz2002} and \citet[][green]{deBoer2017}. The background colours indicate the remnant mass predictions according to \citet{Patton2020}, where dark grey regions correspond to black hole formation.
    }
    \label{fig:all_MCO}
\end{figure}

\subsection{Fusion reactions $\rm ^{12}C+^{12}\!C$, $\rm ^{12}C+^{16}\!O$, and $\rm ^{16}O+^{16}\!O$}

After the formation of the CO core, the fusion reactions of $\rm ^{12}$C and $\rm ^{16}$O will continue to impact its internal structure. The different rates tested for these two fusion reactions lead to variations in the convective core masses during C-, Ne-, and O-burning, as well as in their respective lifetimes (see Tab.~\ref{tab:lifetime}). We also note variations in the convective shell burning, both in the shell mass and in the duration of the burning episode. From the Kippenhahn diagrams of the 20\,M$_{\odot}$ non-rotating models (App.~\ref{sec:Kippenhahn}), we observe the resulting differences in shell C-burning, particularly in the extent of the second carbon shell. In the HIN(RES) model, the convective shell extends almost 0.5\,M$_{\odot}$ further out compared to TDHF, thereby altering the compactness evolution in this layer (see App.~\ref{sec:Kippenhahn}). 
 
The contraction of the models during the second carbon shell episode is faster than during the first. The smaller C-burning shell of the THDF framework displays a comparatively shallow contraction relative to, e.g.  the HIN(RES) shell (noticeable by the smaller slope of the iso-radii lines in Fig.~\ref{fig:kippen}).
The contraction rate differs significantly among models during the Ne-burning phase. A longer Ne-burning phase sustains the contraction of the outer shells longer. The TDHF calculations yield the shortest Ne-burning phase, resulting in a correspondingly brief contraction period of the outer layers.
\begin{figure}[t]
    \center
    \includegraphics[width=90mm]{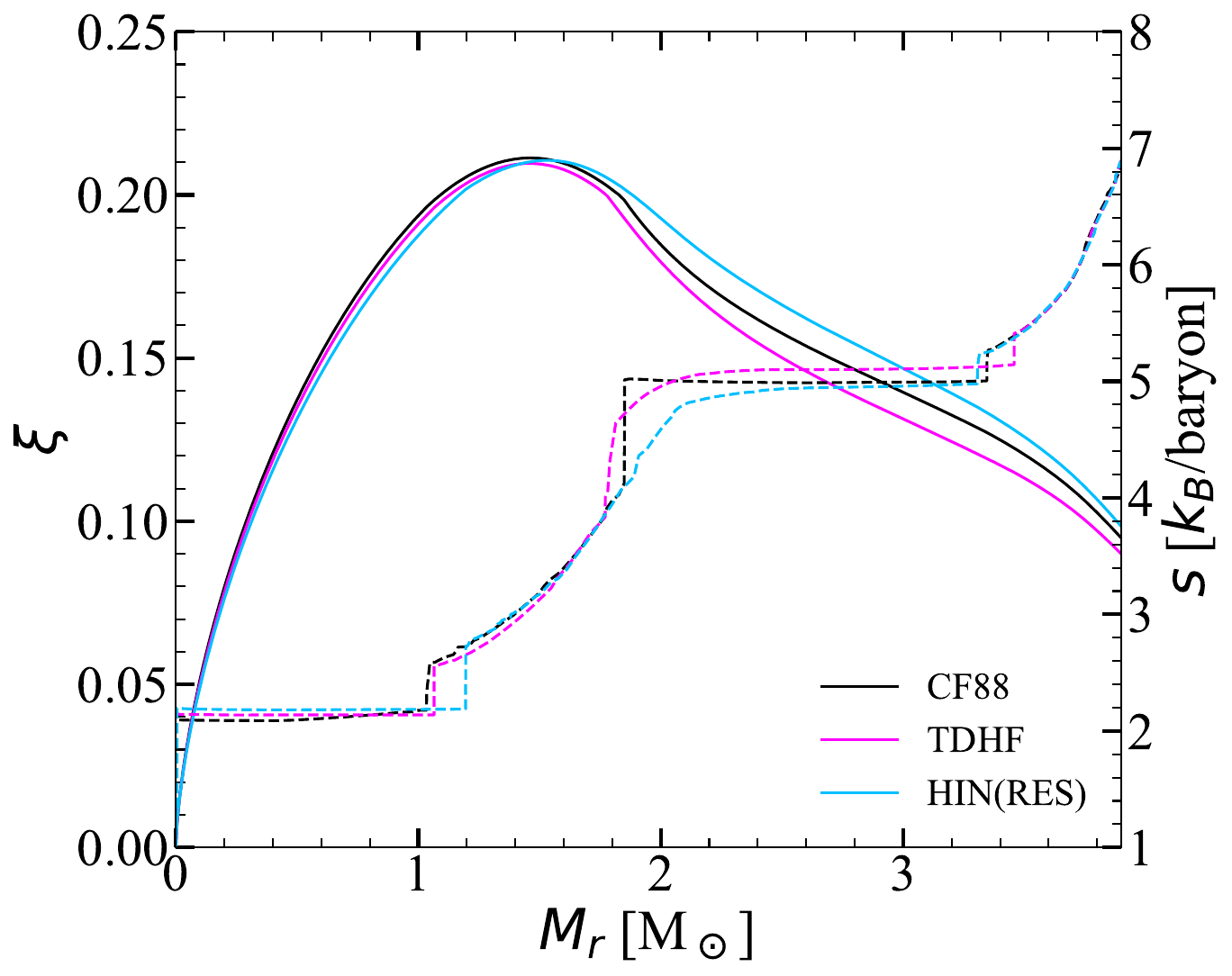}
    \caption{Compactness (full line), defined in Eq.~\ref{eq:xi}, and entropy (dashed line) for the 20\,M$_{\odot}$ models using the rates from CF88, TDHF, and HIN(RES). Models are shown at the midway through core O-burning. 
    }
    \label{fig:xi_s.pdf}
\end{figure}

The results of these differences in core and shell burning shape the compactness profile and the entropy distribution within the CO core at late times. In Fig.\ref{fig:xi_s.pdf}, we show the distribution of entropy and compactness as a function of Lagrangian mass near the end of core O-burning for the same models as in App.~\ref{sec:Kippenhahn}. We note that the peak compactness, reached at about 1.4\,M$_{\odot}$, is similar across all models. More importantly, the TDHF model shows the lowest compactness between $2\lesssim M_r \lesssim 3.5\,$M$_{\odot}$. This corresponds precisely to the region affected by the second C-burning shell, whose extent and duration vary significantly among the TDHF, CF88, and HIN(RES) scenarios. While compactness is commonly used as a proxy for explosiveness \citep[][but see \citealt{Aloy_2021MNRAS.500.4365,Wang_2022MNRAS.517..543,Burrows_2024}]{O'Connor_2011}, in this study we have assessed the explosion outcome primarily through the CO core mass and the central $\rm ^{12}$C abundance, finding that all three models considered here result in successful explosions and form NSs.  
Although these models have not evolved all the way to collapse, we already observe indications that changes in nuclear reaction rates impact the density and compactness structure, with potential consequences for the explosion mechanism. In Fig.~\ref{fig:xi_s.pdf}, we also represent the entropy per baryon $S$ to illustrate the structural differences among the scenarios. We highlight the variation in shape and location of the entropy discontinuities. The first entropy jump in the HIN(RES) model occurs furthest out due to a larger O-burning core. Additionally, the second entropy jump is smoother in HIN(RES) and sharper in CF88. This reflects the fact that the CF88 model retains an active C-burning shell during O-burning, whereas in the HIN(RES) case, the last convective shell vanishes nearly a year earlier, allowing for diffusive smoothing and contraction in that region. The TDHF cross-sections do not induce an active C-burning shell during the O-burning either. Nevertheless, only a few weeks elapse between the shutdown of O-burning and the time at which the compactness and entropy are evaluated in Fig.~\ref{fig:xi_s.pdf}. Therefore, the entropy discontinuity does not have sufficient time to diffuse, unlike in the HIN(RES) case. The location, shape, and magnitude of these entropy jumps are another critical factor in determining the explosion outcome of massive stars \citep{2025_Imasheva}, providing further evidence of structural variability in the CO core driven by the changes in fusion reaction rates explored in this study. 

\section{Summary and outlook} 
\label{sect:CONCLUSION}
We have computed models of massive stars to explore the sensitivity of four key nuclear reactions from He-burning to O-burning phases. In agreement with previous studies, we confirm that the reactions $\rm ^{12}C(\alpha,\gamma)^{16}\!O$ and $\rm ^{12}C+^{12}\!C$ are among the most influential in shaping the stellar structure during the advanced burning stages. The former reaction, active during He-burning, governs the $^{12}$C/$^{16}$O ratio in the core at He-exhaustion and thus sets the amount of fuel available for the subsequent C-burning phase, which is governed by the latter. C-burning is the longest of advanced burning stages, during which the energy produced in the core is mainly evacuated by neutrinos. A longer C-burning phase implies greater energy and entropy losses, thereby increasing the core's susceptibility to degeneracy effects. We confirm that lower $\rm ^{12}C(\alpha,\gamma)^{16}\!O$ rates result in more residual carbon at He-exhaustion and to the formation of less massive cores, delaying the onset of core collapse. In our models, the reduced rates from \citet{deBoer2017} result in shorter He-burning phases, longer C-burning phases, high $^{12}$C/$^{16}$O ratios, and slightly smaller CO core masses (Fig.~\ref{fig:all_MCO}). Consequently, the final fate of the star is also impacted by the adopted rates. Using the remnant prediction scheme by \citet{Patton2020} (developed for non-rotating models), we demonstrate that the differences in nuclear reaction rates can alter the remnant mass and type of non-rotating models. For example, our 20\,M$_{\odot}$ model is predicted to form a BH when using the rates from \citet{Kunz2002}, but an NS when using those from \citet{deBoer2017}.

Variations in the $\rm ^{12}C+^{12}\!C$ and $\rm ^{16}O+^{16}\!O$ fusion rates primarily affect the lifetimes and ignition conditions of C- and O-burning phases, which in turn influence the cores susceptibility to degeneracy effects. Under the HIN(RES) scenario, these rates lead to shorter lifetimes and ignition at higher temperatures and densities, whereas TDHF rates yield longer C-burning lifetimes and ignition at lower temperatures and densities, albeit to a lesser extent. These differences alter both the abundances of the major isotopes and the internal structure. In particular, the extent of the convective C-burning shells responds to the adopted rates: they expand under lower fusion probabilities and contract under higher ones, thereby modifying the mass distribution and the compactness profile. Although the updated $\rm ^{12}C+^{12}\!C$ and $\rm ^{16}O+^{16}\!O$ rates induce relatively modest differences in the global nucleosynthetic yields, they produce significant changes in the chemical structure during advanced combustion stages. These structural differences are expected to impact the final ejecta composition, motivating future dedicated studies. Regarding the $\rm ^{12}C+^{16}\!O$ reaction, we find that only changes in the rates by several orders of magnitude ($\approx 10^4$) would noticeably alter the nucleosynthesis during the shell C-burning phase, where this reaction should operate. The two new rate models explored in this work do not reach such deviations and are therefore not expected to significantly influence the associated nucleosynthesis. 

On the one hand, experimental measurements of fusion reactions at astrophysical energies are challenged by the extremely low cross-sections involved. On the other hand, theoretical approaches, such as the TDHF method used in this work, can estimate reaction rates at these low energies, but with the limitation that they do not account for resonances in the compound system or its nuclear level density, particularly relevant for $\rm ^{12}C+^{12}\!C$ fusion. These two approaches are thus complementary and both are essential for improving stellar evolution models. Progress in both experimental methods and theoretical frameworks will be critical to achieve precise and unified reaction rates inputs for stellar modelling. In this context, a recent advancement in TDHF modelling has demonstrated its ability to reproduce certain resonant structures of the $\rm ^{12}C+^{12}\!C$ system, offering promising new insights on them \citep[][]{Close2025}. Finally, it is important to stress that the nuclear reaction rates investigated here represent only one component of a much broader and complex picture. Our results also depend sensitively on the treatment of mixing processes. Uncertainties in angular momentum transport, convective boundary mixing, or mass loss through stellar winds can all influence the outcomes, in some cases more significantly than variations in nuclear rates. Particularly, the prescriptions adopted for stellar rotation and the possible role of magnetic fields exert cumulative effects throughout the stellar evolution, ultimately shaping its final fate \citep[e.g.][]{Meynet2013,Dumont2021a,Nandal2023}. These aspects will be the subject of a forthcoming study (Dumont et al. in prep.). 

\begin{acknowledgements}
This work was supported by the European Union (ChETEC-INFRA, project no. 101008324). We thank the anonymous referee for helpful comments on the manuscript. T.D. thanks S. Ekström and R. Hirschi for helpful discussions. MAA and AG acknowledge the support through the grants PID2021-127495NB-I00 and PID2021-125485NB-C21 funded by MCIN/AEI/10.13039/501100011033 and by the European Union, and the Astrophysics and High Energy Physics programme of the Generalitat Valenciana ASFAE/2022/026 funded by MCIN and the European Union NextGenerationEU (PRTR-C17.I1) as well as support from the Prometeo excellence programme grant CIPROM/2022/13 funded by the Generalitat Valenciana. A.C. is a Postdoctoral Researcher of the Fonds de la Recherche Scientifique – FNRS. K.G. acknowledges support from DE-SC0023175 (Office of Science, NUCLEI SciDAC-5 collaboration). This research has made use of NASA's Astrophysics Data System Bibliographic Services.
\end{acknowledgements}

%
%
\bibliographystyle{aa}
\bibliography{references.bib}

\begin{appendix}

\section{Nuclear network}
\label{Annexe:nucl_net}
The full set of isotopes included in the GenValNet48 network of the GENEC code is listed in Tab.~\ref{tab:network}.
\begin{table}[h]
    \centering
     \caption{Isotopes included in the 48 species network of GENEC.}
\begin{tabular}{ll|ll}
\hline
\hline
\textbf{Element} & $\rm A_{values}$  & \textbf{Element} & $\rm A_{values}$ \\
\hline
$n\phantom{\displaystyle 1\over 1}$  & 1    & P    & 31  \\
H    & 1    & S   & 32 ; 34 \\
He   & 3 ; 4   & Cl  & 35   \\
 C    & 12 ; 13 ; 14   & Ar  & 36 ; 38  \\
 N    & 14 ; 15    & K   & 39   \\
O    & 16 ; 17 ; 18  & Ca  & 40 ; 42  \\
F    & 19  & Ti  & 44 ; 46\\
Ne   & 20 ; 21 ; 22  & Cr  & 48 ; 50 ; 56  \\
Na   & 23   & Fe  & 52 ; 53 ; 54 ; 55 ; 56 \\
Mg   & 24 ; 25 ; 26 &  Co  & 55 ; 56 ; 57 \\
Al   & 26 ; 27  &  Ni  & 56  \\
Si   & 28 ; 30  &  \\
\hline
\end{tabular}
    \label{tab:network}
    \tablefoot{$\rm A_{values}$ are the discrete values of the mass number for each species that is included in the network.}
\end{table}

\section{Nuclear reaction rates}
\label{Annexe0}

\subsection{$\rm ^{12}C+^{12}\!C$ theoretical predictions}
\label{sec:c12c12}
As discussed in Sect.\,\ref{subsubsection:tdhf}, several energy density functionals have been used in the TDHF calculations for the $\rm ^{12}C+^{12}\!C$ nuclear reaction. Figure~\ref{fig:ratesall} presents the resulting S-factor predictions obtained using three of these functionals. In the astrophysical energy range of interest, all TDHF predictions lie above those from the other models compared in this work. Since no specific functional is favoured by the authors, we adopt the results from the Skyrme SLy4d functional (shown in red), which gives the most pronounced deviations, for our sensitivity analysis.

\begin{figure}[hbt!]
    \center
    \includegraphics[width=90mm,trim=0mm 0mm 0mm 0mm,clip]{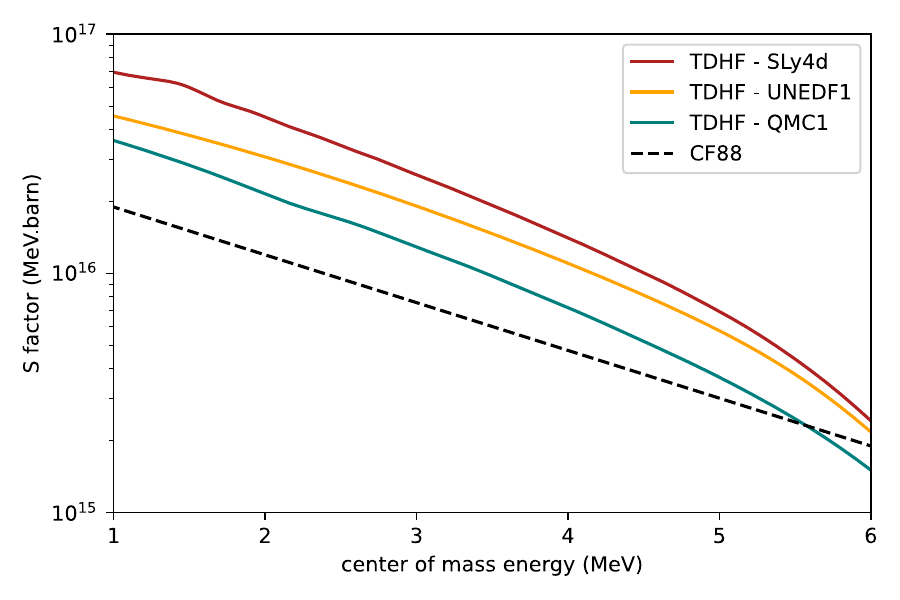}
    \caption{S-factor model predictions for the $\rm ^{12}C+^{12}\!C$ fusion reaction as a function of the relative energy E$\rm _{rel}$, obtained using different functionals in the TDHF calculations. The SLy4d functional (in red) is the one considered in this work. For comparison, the CF88 reference is indicated by the black dotted line.}
    \label{fig:ratesall}
\end{figure}

\subsection{Neutron source reactions}
The two primary nuclear reactions driving the slow neutron-capture process are $\rm ^{13}C(\alpha,n)^{16}O$ and $\rm ^{22}Ne(\alpha,n)^{25}Mg$. In rotating stellar models, the $\rm ^{13}C(\alpha,n)^{16}O$ reaction is favoured due to the higher temperatures reached in the core and in the burning shells. 
In the present work, we have updated the rates for both reactions. Figure~\ref{fig:rateC13} gives the $\rm ^{13}C(\alpha,n)^{16}O$ reaction rates adopted in this study, and based on the evaluation by \citet{Ciani2021}, in comparison to the NACRE compilation used in D24. During the He- and C-burning phases, the updated rates are slightly lower than for the NACRE reference.
 \begin{figure}[hbt!]
    \center
    \includegraphics [width=90mm,trim=10mm 10mm 10mm 5mm,clip]{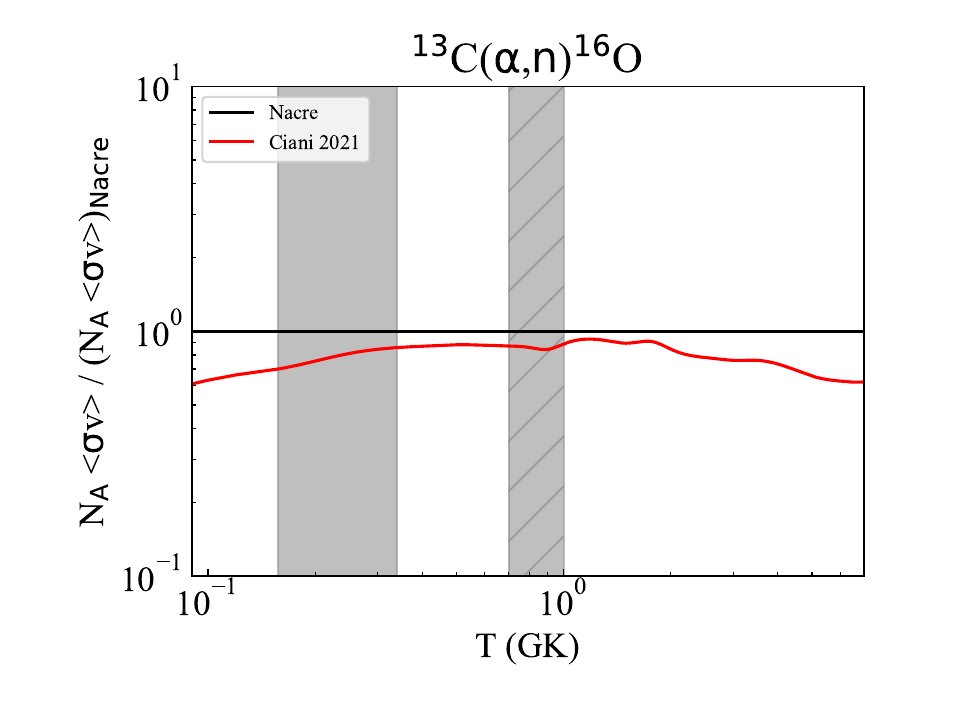}
    \caption{Reaction rates for $\rm ^{13}C(\alpha,n)^{16}O$ from \citet{Ciani2021}, normalised to the NACRE reference rates \citep{Angulo1999}. Dark grey shade areas indicate the temperature ranges corresponding to He- and C-burning in a rotating 17~M$_{\odot}$ GENEC model with an initial rotation rate of $\rm V_{ini}/V_{crit} = 0.4$.}
    \label{fig:rateC13}
 \end{figure}
 \begin{figure}[hbt!]
    \center
    \includegraphics [width=90mm,trim=10mm 10mm 10mm 5mm,clip]{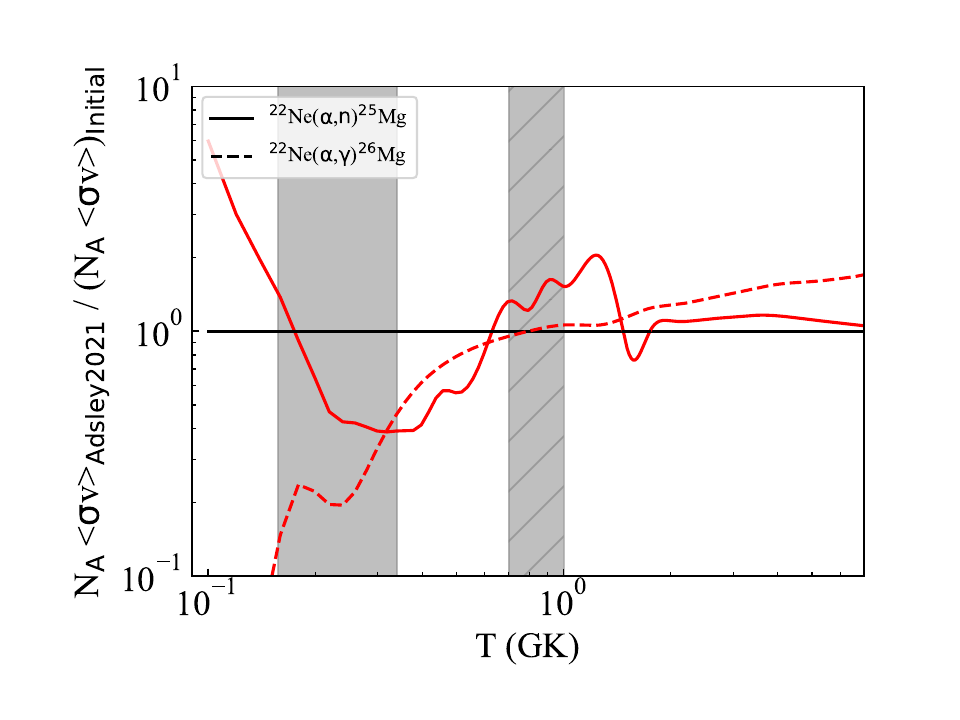}
    \caption{Reaction rates from \citet{Adsley2021}, normalised to the reference rates of \citep{Iliadis2010}, for the $\rm ^{22}Ne(\alpha,n)^{25}Mg$ and $\rm ^{22}Ne(\alpha,\gamma)^{26}Mg$ nuclear reactions. Dark grey shade areas indicate the temperature ranges corresponding to He- and C-burning in a rotating 17~M$_{\odot}$ GENEC model with an initial rotation rate of $\rm V_{ini}/V_{crit} = 0.4$.}
    \label{fig:rateNe22}
 \end{figure}
Figure~\ref{fig:rateNe22} shows the $\rm ^{22}Ne(\alpha,n)^{25}Mg$ reaction rates from \citet{Adsley2021} used in the present work, compared to the reference rates from \citep{Iliadis2010} used in D24. During the He- and C-burning phases, we expect lower and higher reaction rates, respectively. The same figure also displays the reaction rates for the competing $\rm ^{22}Ne(\alpha,\gamma)^{26}Mg$ reaction, which acts in opposition to the neutron-producing channel.

\section{Nucleosynthesis one-zone code: Shell C-burning trajectories}
\label{Annexe3}
\label{sec:nucleosynthesis-trajectories}

To select a representative trajectory for the temperature and density conditions during carbon shell burning, we identify, at each step of the evolution, the location where the energy production from the $^{12}$C+$^{12}$C reaction peaks. This location consistently corresponds to the base of the convective shell burning zone, as illustrated, e.g. in Fig.~\ref{fig:energy_profiles}. In most of the models, two successive convective C-burning shells develop. The evolution of temperature and density within these shells, plotted as a function of the $^{12}$C mass fraction is shown in Fig. \ref{fig:shell_burning_trajectories} for a rotating 15~M$_{\odot}$ star and the three nuclear reaction rates references: CF88, HINRES, and TDHF. The second shell exhibits slightly higher temperatures, while the densities remains comparable across the two shells. It is worth noting that the second shell has a shorter lifetime (about 1/8 shorter), resulting in a correspondingly reduced neutron exposure.

\begin{figure}[hbt!]
    \centering
    \includegraphics[width=\linewidth]{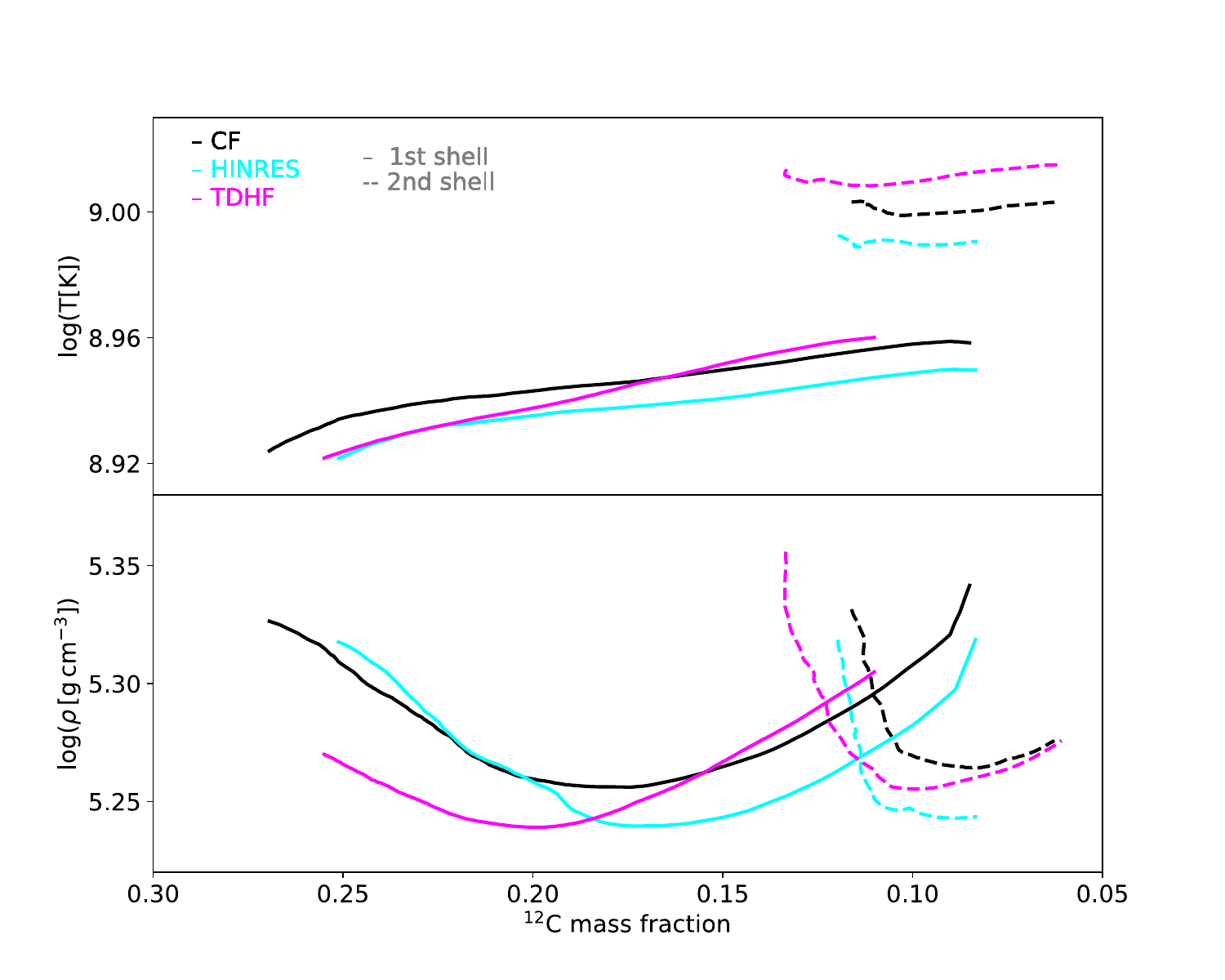}
    \caption{Temperature and density conditions within the first (solid lines) and second (dashed lines) carbon burning shells for the 15~M$_{\odot}$ rotating star, shown as a function of the $^{12}$C mass fraction.}
    \label{fig:shell_burning_trajectories}
\end{figure}

\section{Kippenhahn diagrams}
\label{sec:Kippenhahn}

Kippenhan diagrams are a useful tool to understand the internal evolution of stellar models under different nuclear reaction rate scenarios. 
\begin{figure}[htb!]
    {\includegraphics[width=82mm,trim=45mm 30mm 40mm 10mm, clip]{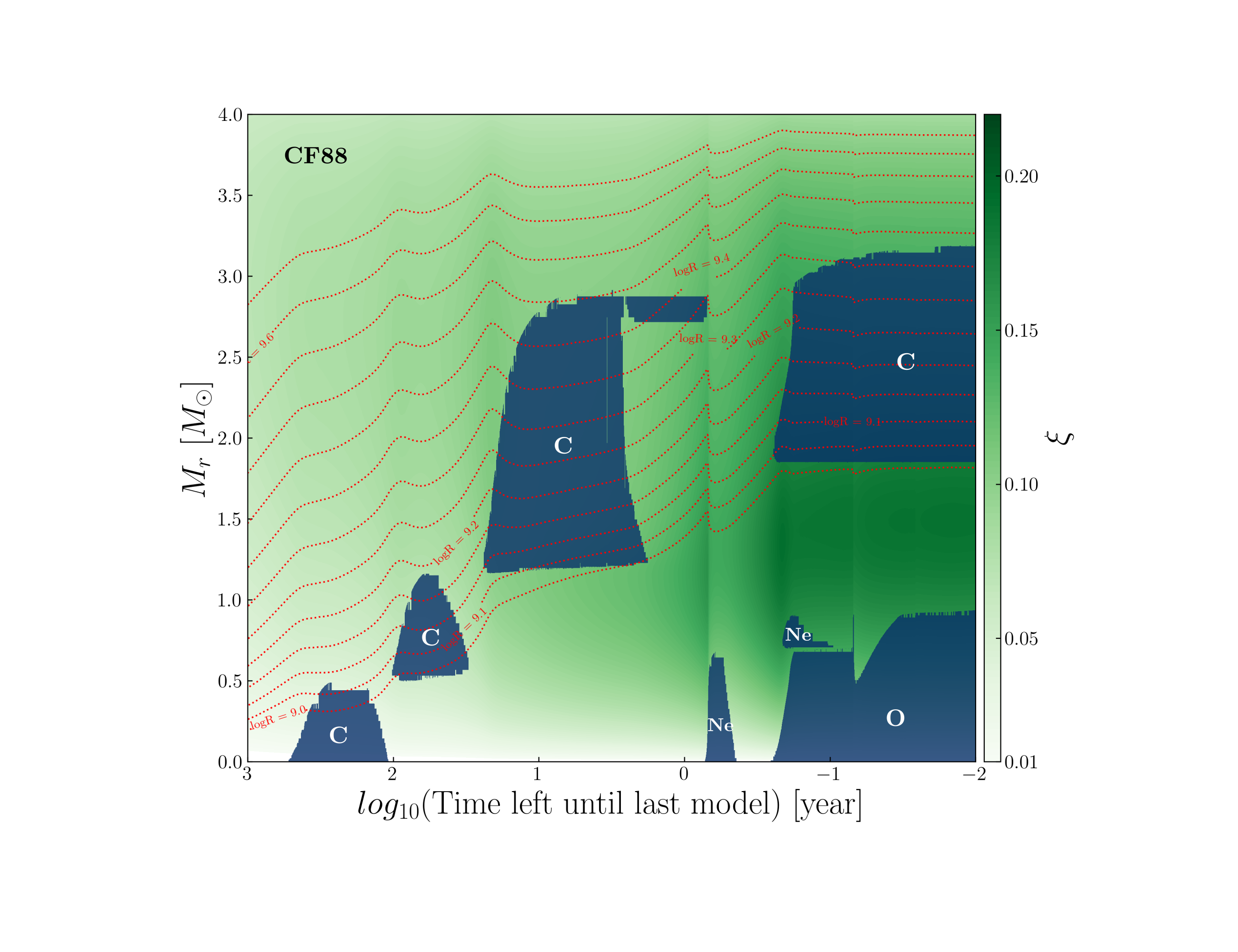}}\\
    {\includegraphics[width=82mm,trim=45mm 30mm 40mm 20mm, clip]{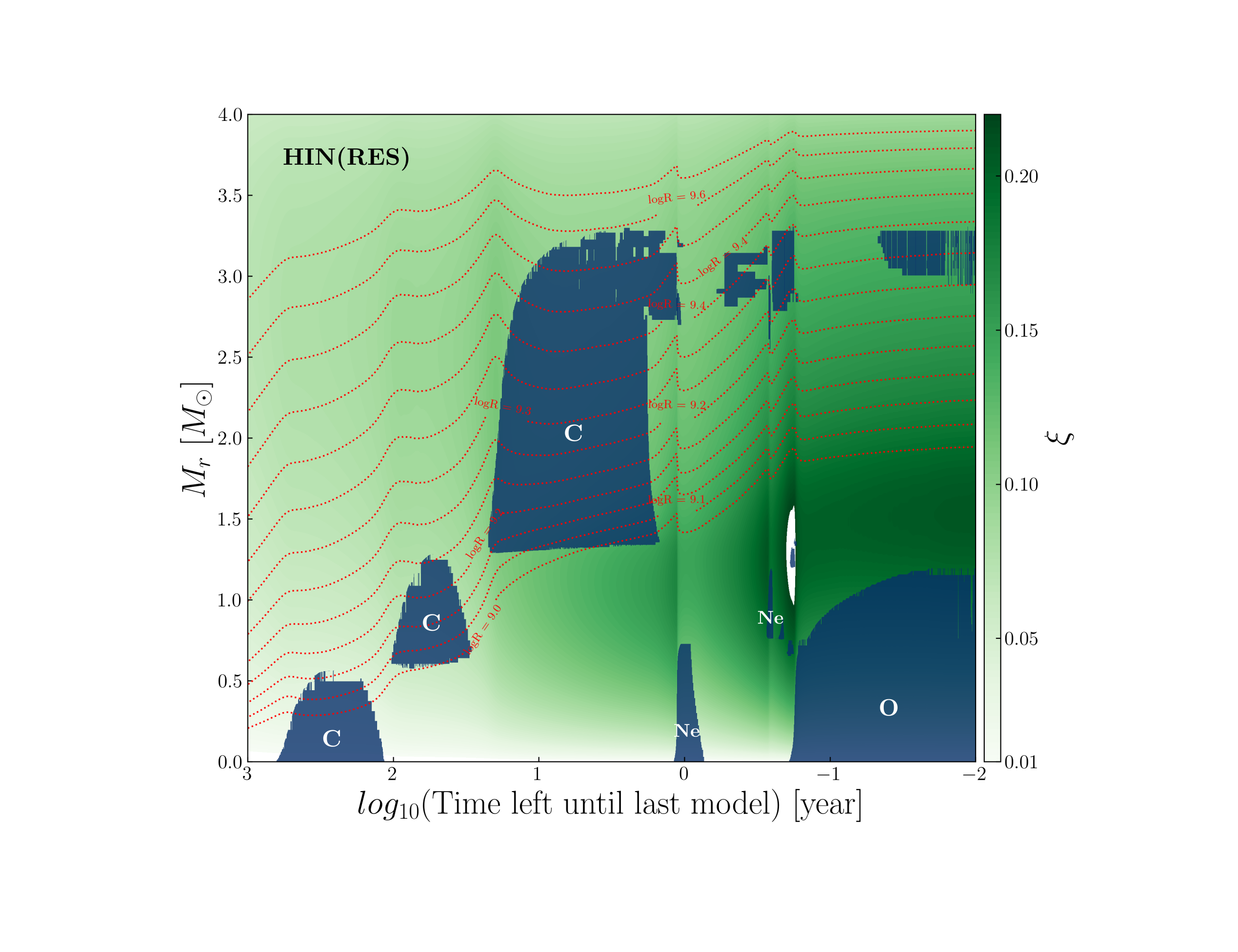}} \\
    \includegraphics[width=82mm,trim=45mm 30mm 40mm 20mm, clip]{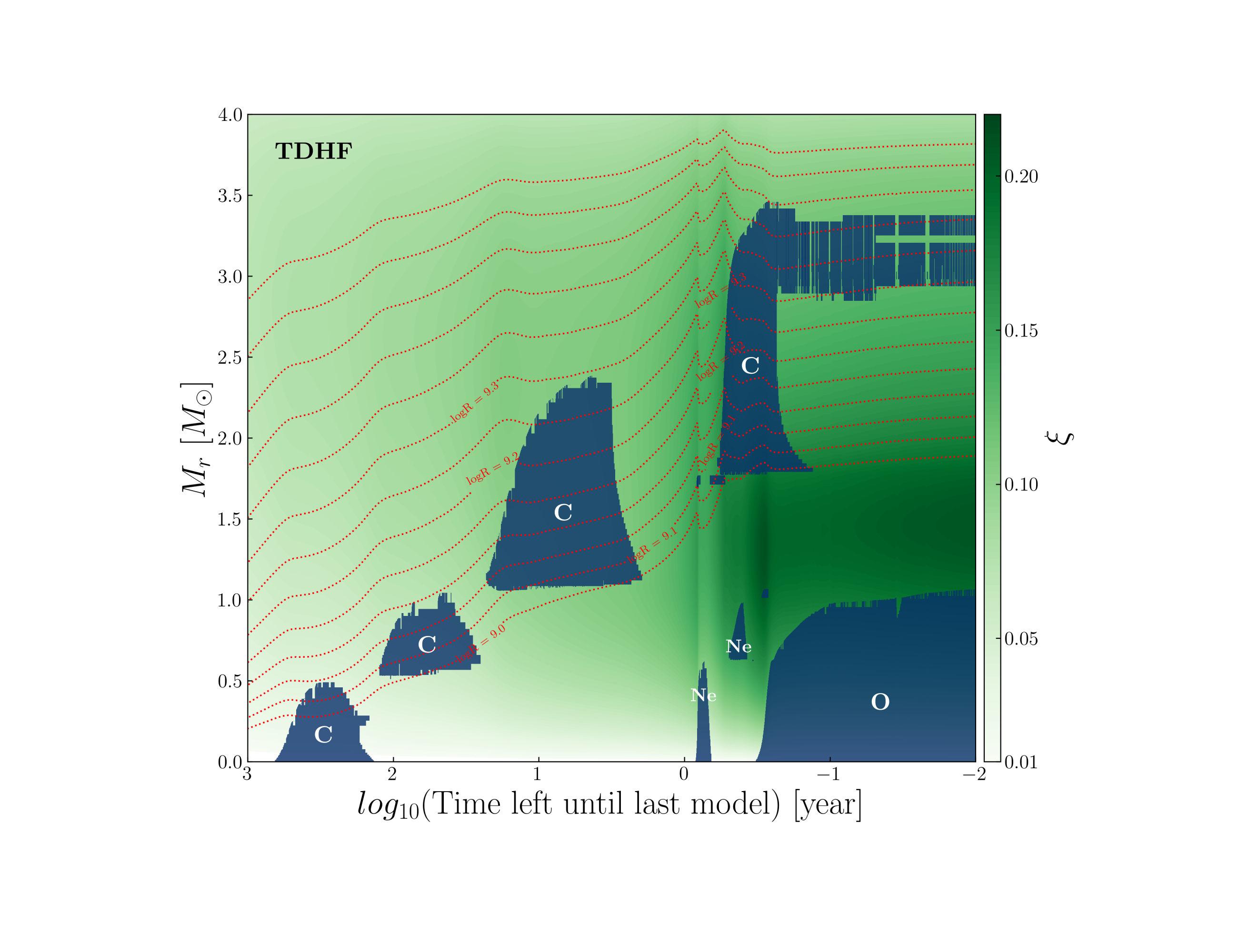}
    \caption{Kippenhahn diagrams of the 20\,M$_{\odot}$ non-rotating stars from the core C-burning to the core O-burning for the three nuclear reaction rate scenarios. Dark blue shade shows the convective zones. Iso-radii contours are shown by dotted red lines. The compactness $\xi$ is indicated by a colour bar. Top: CF88. Middle: HIN(RES). Bottom: TDHF. %
    }
    \label{fig:kippen}
 \end{figure} 
Figure~\ref{fig:kippen} shows the Kippenhahn diagrams for the 20~M$_{\odot}$ non-rotating models explored in this study.
Each panel illustrates the convective zones during the core C-burning phase, the successive shell C-burning episodes, core and shell Ne-burning, and core O-burning. A significant effect of the nuclear rates variations is observed in the extent of the C-burning shells. Especially, the second C-burning shell is significantly more compact in the TDHF model compared to the CF88 and HIN(RES) cases. 

In the Kippenhahn diagrams, the background is colour-coded to reflect the compactness of each mass shell, defined as
\begin{align}
\label{eq:xi}
\xi = \frac{M_r/M_\odot}{r/1000\,\text{km}}.
\end{align}
An overall trend of increasing compactness is evident by the darker green shades, especially towards the end of the computed evolution and in the vicinity of the convective O-burning core.
To illustrate the contraction of the selected mass shells, we also overlay iso-radius contours in Fig.~\ref{fig:kippen}. The contraction manifests as increasing mass accumulated at each considered radius (hence, iso-radius contours grow secularly).

\section{Remnant properties}
\label{sec:remmnant_properties}
Table~\ref{tab:Xc_MCO} presents the remnant properties predicted by our non-rotating models. The CO core mass is determined at the end of the He-burning phase and is defined as the mass coordinate where the central helium mass fraction drops below $\rm X_{He}<10^{-1}$. In cases where the model predicts black hole formation, the remnant mass corresponds to that of the He core, defined as the mass coordinate where the central hydrogen mass fraction falls below $\rm X_{H}<10^{-1}$. The remnant type is inferred based on the values of $\rm X_{^{12}C}$ and $\rm M_{CO}$ following the criteria of \citet{Patton2020}. For models resulting in NS formation, the remnant mass $\rm M_{rem}$ is taken from the tables of \citet{Patton2020}, which define the NS mass as $\rm M_4$, corresponding to the mass coordinate where the entropy per baryon reaches $4 \rm k_B/baryon$. The last column displays $\rm M_4$ for models that have reached the end of O-burning; this value closely approximates the final mass expected at the pre-SN stage. For models that have not reached the end of O-burning, this quantity is not available and is indicated by ''/''.

\begin{table*}[!hb]
    \centering
    \caption{Remnant properties: type and mass. 
    }
    \begin{tabular}{c|c|c|c|c|c|c|c}
        \multicolumn{7}{c}{} \\
         \hline \hline
         M$\rm _{ini}$  & Rates  & $X_{^{12}C}$ & $\rm M_{CO}$ (M$_{\odot}$) & REM & $\rm M_{rem}$ (M$_{\odot}$) &  $\rm M_4/M_{CO}$ & M$\rm _{Fin}$ (M$_{\odot}$) \\
         \hline
         15~M$_{\odot}$ &  K02 + CF88&  0.349 & $2.49^{\textcolor{red}{*}}$ & NS & 1.41  & / & 13.9 \\
         15~M$_{\odot}$ &  DB17 + CF88&  0.387 & $2.47^{\textcolor{red}{*}}$ &  NS & 1.46 &  0.66 & 13.9 \\
         17~M$_{\odot}$ &  K02 + CF88&  0.331 & 3.06& NS & 1.62  & / & 15.8 \\
         17~M$_{\odot}$ &  DB17 + CF88&  0.366 & 3.04 &  NS & 1.61 &  0.54 & 16.0 \\
         20~M$_{\odot}$ &  K02 + CF88&  0.296 & 4.27& BH & 10.3 & /  & 10.5 \\
         20~M$_{\odot}$ &  DB17 + CF88&  0.344 & 4.10 &  NS & 1.51 & 0.43 & 10.9 \\
         \hline
         \hline
        15~M$_{\odot}$ &  DB17 + HINRES&  0.387  & $2.47^{\textcolor{red}{*}}$ & NS &  1.46  & 0.63 & 13.9 \\
        15~M$_{\odot}$ &  DB17 + TDHF& 0.387  &  $2.48^{\textcolor{red}{*}}$ & NS &  1.46 & 0.65 & 13.9 \\
        17~M$_{\odot}$ &  DB17 + HINRES& 0.366   & 3.05 & NS &  1.60 & 0.55 & 16.0 \\
        17~M$_{\odot}$ &  DB17 + TDHF&  0.366  &  3.03 & NS &  1.61 & 0.53 & 16.0 \\
        20~M$_{\odot}$ &  DB17 + HINRES& 0.346   & 4.08 & NS  &  1.49 & 0.43 & 10.9 \\
        20~M$_{\odot}$ &  DB17 + TDHF& 0.346   & 4.09  & NS &  1.49& 0.44 & 10.9  \\
         \hline
         \hline
    \end{tabular}
    \label{tab:Xc_MCO}
    \tablefoot{Remnant properties are inferred from the values of the $^{12}$C mass fraction and the CO core mass, both evaluated at the end of core O-burning (or at the end of core C-burning if O-burning is not finished), following the prescription of \citet{Patton2020}. Initial mass (Col. 1), rates reference where K02 and DB17 refer to \citet{Kunz2002} and \citet{deBoer2017}, respectively (Col. 2), $^{12}$C mass fraction (Col. 3), CO core mass (Col. 4), remnant type where NS and BH refer to neutron star and black hole, respectively (Col. 5), remnant mass (Col. 6), $\rm M_4$ to CO core mass ratio (Col. 7, models marked with $^{\textcolor{red}{*}}$ have a CO core mass outside the \citet{Patton2020} grid, whose minimum of 2.5 M$_{\odot}$, predictions for these cases rely on a small extrapolation and final mass (Col. 8). We note that M$_4$ is only computed at the end of O-burning, but using the results of \citet{Griffiths2025} we checked that this value is at most 5\% off from the value at pre-SN. We expect this to hold true here also. 
    }
\end{table*}

\label{LastPage}
\end{appendix}

\end{document}